\documentclass[english,a4paper]{article}

\usepackage{amsmath}
\usepackage{amsfonts}
\usepackage{amssymb}

\usepackage{authblk}
\usepackage{hyperref}
\usepackage[english]{babel}
\usepackage[numbers]{natbib}

\newcommand{\textcite}[1]{\cite{#1}}
\renewcommand{\cite}[1]{\citep{#1}}

\usepackage{xcolor}
\usepackage{hyperref}
\usepackage{listings}
\usepackage{cleveref}
\usepackage{subcaption}

\newcounter{myOctave}
\newcommand{\myOctave}{\noindent \refstepcounter{myOctave}\texttt{octave:\arabic{myOctave}> }}

\newcommand{\CC}{C\nolinebreak\hspace{-.05em}\raisebox{.4ex}{\scriptsize\bf +}\nolinebreak\hspace{-.10em}\raisebox{.4ex}{\scriptsize\bf +}}

\usepackage{xfrac}
\usepackage{multicol}

\usepackage{threeparttable,booktabs,tabularx}
\usepackage{nolbreaks}

\usepackage{epigraph}

\usepackage{url}

\title{Mirror Matrix on the Wall: coding and vector notation as tools for introspection}

\author[1]{Leonardo Araújo\thanks{Email: \href{mailto:leolca@ufsj.edu.br}{leolca@ufsj.edu.br}}}
\affil[1]{Universidade Federal de São João del Rei, DTECH, Ouro Branco, MG, Brasil.}

\providecommand{\keywords}[1]{\textbf{\textit{keywords:}} #1}

\begin{document}
\maketitle

\begin{abstract}
    The vector notation adopted by GNU Octave plays a significant role as a tool
for introspection, aligning itself with the vision of Kenneth E. Iverson. He
believed that, just like mathematics, a programming language should be an
effective thinking tool for representing and reasoning about problems we wish
to address. This work aims to explore the use of vector notation in GNU Octave
through the analysis of operators and functions, providing a closer alignment
with mathematical notation and enhancing code efficiency. We will delve into
fundamental concepts such as indexing, broadcasting, and function handles, and
present case studies for a deeper understanding of these concepts. By adopting
vector notation, GNU Octave becomes a powerful tool for mathematicians,
scientists and engineers, enabling them to express and solve complex problems
more effectively and intuitively.
 
\end{abstract}

\section{Introduction}\label{sec-intro}

\epigraph{One reason, or rather one explanation of why languages differ is that speech communities differ with respect to the ways they use their language(s) and the communicative functions that languages fulfill}{Florian Coulmas, \emph{Language adaptation}}

A programming language is a formal system used to write computer programs or
provide instructions to a computer \cite{abelson1996}. Programs aim to solve a
problem, perform a simulation, task, or function. The applications can vary
widely, ranging from text editing and musical composition to launch and
trajectory control a rocket. Each of these distinct tasks can be approached
through various representations. Programming languages use syntax and semantics
to create abstractions, through which they define and manipulate data and
structures, as well as control the flow of execution. Certain programming
languages are more suited to handle specific types of problems compared to
others, leading to simpler notation and solutions, greater efficiency, and
enhance our ability to think, communicate, and solve more effectively
\cite{mcconnell2004}.

Computational tools can be used to complement mathematical thinking in various
ways. First, as a means of conceptualization and reasoning, example creation or
counterexample search, data creation or simulations that could help suggest
conjectures, and even assist in proofs or theorems theorems\footnote{The
    four-color theorem was the first theorem proved using computation. Kenneth
    Appel and Wolfgang Haken reduced the problem to a finite (though large)
    number of cases, which were tested computationally. Computers have since
    been used in various other mathematical proofs, such as Fermat's Last 
    Theorem and Kepler's conjecture, among others} \cite{dubinsky2002}.

\begin{quote} 
  ``Indeed, it is in general true that whenever a person constructs
  something on a computer, a corresponding construction is made in the
  person's mind. It is possible to orchestrate this correspondence by
  providing programming tasks in an appropriate programming language designed
  so that the resulting mental constructions are powerful ideas that enhance the
  student's mathematical knowledge and understanding''.\cite{dubinsky2002}
\end{quote}

Just as programming offers different ways to represent and solve problems,
introspection--the act of looking inward--relies on a clear and structured
expression of our ideas.  The concept of `Mirror Matrix on the Wall', inspired
by the famous phrase \emph{mirror, mirror on the wall}, refers to
introspection, where the mirror serves as a tool for internal reflection. In
the classic \emph{Snow White} story, the mirror answers the queen's questions,
revealing hidden truths. Similarly, each programming approach acts as a
different `mirror' for thought, reflecting in distinct ways and creating
various abstractions for solving complex problems. Choosing the right notation
for each problem allows for the development of more elegant and efficient
solutions.

In undergraduate courses, procedural paradigms are commonly used as an
introductory approach to the programming world\footnote{In the study conducted
  by \textcite{deraadt2004}, instructors reported using one of the following
  paradigms in introductory programming courses: procedural, object-oriented,
  and functional. In most cases, the procedural approach was adopted, even when
  using an object-oriented programming language. Similar findings were reported
  by \textcite{mason2024}.} \cite{deraadt2004,mason2024}. Procedures are outlined
by a series of computational steps that must be followed to perform specific
tasks. This approach aims to simplify the understanding of programming concepts
for beginners. The adopted concepts are more accessible, easier to assimilate,
and require a lower level of abstraction from students. The structural
approach plays a fundamental role in conveying the principles of code
organization through functions and procedures. Procedural languages also
facilitate learning by introducing basic programming elements, such as
variables, data types, control flow structures, and functions. The inherent
sequential nature of these languages resembles the sequential thinking often
used in problem-solving, fostering logical reasoning in algorithmic
construction. The procedural approach also demands less abstraction.
Apparently, the combination of all these factors makes it a more accessible and
straightforward programming paradigm, which is why it is predominantly adopted
in introductory programming courses. However, some studies show that the
object-oriented paradigm can be used in introductory courses, yielding results
similar to those observed when the procedural approach is used
\cite{wiedenbeck1999,vilner2007fundamental}.

The intersection between programming languages and mathematical notation must
be considered, especially when dealing with problems that are modeled and
solved mathematically. As Iverson \cite{iverson1980} expressed in his work, ``most
programming languages are decidedly inferior to mathematical notation and are
little used as tools of thought in ways that would be considered
significant'' \cite{iverson1980}.
This provocative statement leads us to explore how traditional programming
languages contrast with the effectiveness of mathematical notation as a tool
for expression and reasoning.

In this context, the APL programming language\footnote{Named after the book
\emph{A Programming Language}.}, developed by Iverson in the 1960s, stands out
for introducing an approach aligned with the concept of \emph{Notation as a Tool of
Thought}. APL utilized multidimensional arrays, operators, and functions, which
led to the development of concise code. In addition to vectorization, APL
pioneered providing an interactive programming environment, allowing for
simultaneous code conception and result analysis \cite{mcleod}. Programming
languages such as APL are aimed at scientific computing and engineering, and use 
the vector programming paradigm. The basic data structures are vectors, taking
advantage of the fact that their constituent elements are of the same or
similar types.

The vector paradigm is used to generalize and apply operators to scalars,
vectors, matrices, or higher-dimensional arrays. This approach offers a higher
level of abstraction, providing a more concise and efficient notation for
representing mathematical problems, allowing the programmer to think and
operate on data structures instead of worrying about explicitly looping over
the scalars that make up these structures. The vector approach is also
inherently suited for parallelism, as operations are performed element by
element. Furthermore, modern vector programming languages can leverage
multicore processors and GPUs for faster computation\footnote{Vector
programming, which involves operations on arrays or matrices of data, can
benefit significantly from SIMD (Single Instruction, Multiple Data)
instructions, such as those found in the SSE (Streaming SIMD Extensions)
families, and from using GPUs (Graphics Processing Units) through
languages like NVIDIA's CUDA (Compute Unified Device Architecture) or
OpenCL (Open Computing Language), which is an open, cross-platform
standard for parallel programming.} \cite{sawadsitang2012}.

A programming language, as a system of notation, must be concise and efficient,
prioritizing the communication of essential information while minimizing the
inclusion of unnecessary details. To achieve this, a compact notation and
higher-level representation should be employed. It is from this perspective
that Iverson argues in favor of using language as a tool for
thought: ``A fundamental scheme for achieving this is the introduction of
grammatical rules by which meaningful phrases and sentences can be constructed
by combining elements of the vocabulary'' \cite{iverson1980}.

Programming languages provide the necessary abstraction for developing complex
systems. This abstraction is achieved by suppressing details, allowing for the
reuse of components and tools, making the complexity of expressing any program
directly in machine language more manageable. A programming language should
enhance the following essential aspects of programs: reading, writing,
execution, and reasoning \cite{orchard2011four}. These principles align with
Iverson's view that an effective and concise notation is essential in a
programming language.

The characteristics of a programming language also connect to the Sapir-Whorf
hypothesis\footnote{According to the Sapir-Whorf hypothesis, language shapes
and determines the cognition and worldview of its speakers.}, suggesting that a
programming language can either facilitate or hinder the representation of
certain concepts and the reasoning about them.

In the comparison below, note how expressing a piecewise function through a
programming language is more intuitive than its mathematical representation:
\begin{center}
\begin{minipage}[m]{0.3\textwidth}
    \lstinline{if x > 0, then y = 1; else y = 0.}
\end{minipage}
\hspace{2cm}
\begin{minipage}[m]{0.3\textwidth}
    \begin{equation*}
    y =
    \begin{cases}
        1 &, x > 0 \\
        0 &, \text{otherwise}.
    \end{cases}
    \end{equation*}
\end{minipage}
\end{center}
Mathematical representation uses the reverse order compared to natural
language, along with conventions, abbreviations, and acronyms. 
Often, the conditional \emph{if} is also omitted, as seen in the first line of the
piecewise definition of $y$.

In this article, we aim to present the use of GNU Octave, highlighting its
characteristics as a vector programming language and demonstrating how to fully
leverage its potential and the advantages of using vector notation. This
approach will enable better structuring, organization, and introspection in
algorithm development, as well as maximize the language's capabilities and
result in significantly more efficient code. Through examples and comparisons,
we also intend to guide the process of transitioning from the procedural
paradigm to the vector paradigm. Clarifying and exposing the vector paradigm is
important, as students in undergraduate courses often finish their studies
using GNU Octave/MATLAB solely within the procedural paradigm.

\section{GNU Octave}\label{sec-octave}

\epigraph{We can only see a short distance ahead, but we can see plenty there
that needs to be done.}{Alan Turing, \emph{Computing machinery and
intelligence}}

GNU Octave is a numerical computing environment that uses a high-level
interpreted programming language. Its main focus lies in mathematical
computations, data analysis and visualization, numerical solutions of linear
and nonlinear problems, as well as algorithm development \cite{octave}.
Offering a free and open-source alternative to MATLAB, GNU Octave is carefully
designed to be compatible with the latter. As a result, it becomes an excellent
option for students and academics seeking a viable alternative without the
costs associated with software licenses.

Octave was initially conceived by John W. Eaton in 1992 \cite{octave}, later
becoming part of the \href{https://www.gnu.org/}{GNU Project}, a free and
open-source software initiative launched by Richard Stallman in 1983. As free
software, GNU Octave receives contributions from a diverse group of
programmers. Like MATLAB, Octave also includes numerous packages. These are
maintained and developed by the community, which can result in a slightly more
limited variety compared to MATLAB. Package contributions are centralized on
\href{https://octave.sourceforge.io/}{Octave Forge}, a collaborative project
dedicated to expanding GNU Octave's functionalities through a comprehensive
selection of packages and tools. This collection covers a wide range of
domains, including signal processing, image processing, optimization,
statistics, control systems, and much more.

Octave is an interpreted language that supports different programming
paradigms, including procedural programming, vector programming, and
object-oriented programming. Additionally, the language is dynamically typed,
meaning there is no need to explicitly declare a variable's type, as it is
automatically determined during program execution. The language supports common
data types such as floating-point numbers, integers, strings, matrices, and
structures. Octave also uses standard control flow structures like loops
(\texttt{for}, \texttt{while}) and conditionals (\texttt{if-else}). Functions
(including anonymous functions) can be defined, and index expressions,
arithmetic operations, comparisons, and assignments can be used. It is also
possible to handle errors and warnings, as well as debug code.
 
\section{Data basic types}

\epigraph{Why, sometimes I've believed as many as six impossible things before
breakfast}{Lewis Carroll, \emph{Through the Looking-Glass, and What Alice Found
There}}

In programming languages, data is internally represented by different types,
each with its own characteristics and limitations. In GNU Octave, the basic
data types can be divided into numeric, strings, logicals, cells, and
structures. While certain vector operations can also be applied to logical
types and strings, the focus here will be on numeric elements. 

The binary
representation of numbers can produce counterintuitive results due to its limited
precision. In repetitive operations, small precision errors can accumulate and
cause significant deviations in the final result. For example, adding
one-thousandth a thousand times may not result exactly in 1 due to these
limitations. Additionally, operations with very large numbers can lead to
overflow\footnote{An overflow occurs when an arithmetic operation produces a
result outside the representable range of the numeric type used.} (or
saturation, if there are mechanisms to prevent it), and when dealing with very
small numbers, we may encounter underflow\footnote{An underflow occurs when an
arithmetic operation results in a number so small that the precision of the
adopted representation is insufficient to represent it correctly.}.

The standard numeric type is \emph{double}, which refers to double precision in
floating-point format. This is a 64-bit representation format defined by the
IEEE 754 standard \cite{ieee754}.
This is adopted as a standard to maintain greater accuracy in numerical calculations 
performed computationally and also because they can represent a wide range of values, 
from very large values to very small ones, without significant loss of precision.

Despite being a good way to computationally represent numbers and perform
calculations, \Cref{lst-binrep} presents an example of how the inherent
limitations of binary floating-point representation can lead to unexpected
results\footnote{To circumvent the limitations imposed by floating-point
arithmetic, there are libraries available for performing arbitrary precision
arithmetic (the term \emph{bignums} is also used to refer to such
representation), limited only by the available memory in the system.}. We must
consider that we are susceptible to errors in the binary representations of
numbers. In this example, the difference between the expected value and the
calculated value is greater than the machine epsilon (\lstinline{eps}).

\begin{lstlisting}[language=octave, escapechar=|, label=lst-binrep, caption={The inherent imprecision of binary representation can lead to unexpected results. Note how the sum of two fractions \sfrac{1}{10} and \sfrac{2}{10} results in a value different from the expected \sfrac{3}{10}.}]
|\myOctave|1/10 + 2/10 == 3/10
ans = 0
|\myOctave|eps
ans = 2.2204e-16
\end{lstlisting}

We are subject to representational errors inherent to binary representation
with finite precision, as well as errors arising from the imprecision of
computational arithmetic calculations. See \Cref{lst-binarith} for an example
of how, even when using well-represented numbers in binary form, we can
encounter inaccuracies stemming from the numerical operation.
\begin{lstlisting}[language=octave, escapechar=|, label=lst-binarith, caption={Imprecision arising from the numerical implementation of division.}]
|\myOctave|a = 0.7777777777777;   
|\myOctave|b = 7;                 
|\myOctave|c = 0.1111111111111;   
|\myOctave|a/b == c
ans = 0
\end{lstlisting}

When defining a variable by assigning it a numeric value, a matrix variable
will be created with elements of type \lstinline{double}. Even if the intention
is to represent a scalar, it will still be represented as a matrix, albeit with
dimensions $1\times 1$. Any variable with a numeric value in GNU Octave will
thus be represented as a matrix. Some numerical operations are integrated
through functions coded in \CC{} and compiled, providing performance and
efficiency in performing numerical calculations.

Among these, we can mention basic arithmetic operations that are implemented
through functions named after the corresponding operation. For example, the
function \lstinline{plus} is responsible for the matrix addition operation,
with the syntax \lstinline{plus(X1, X2, ...)}; however, it is more practical
and intuitive to use the operator overloading form\footnote{Operator
overloading is a specific type of polymorphism where operators have different
implementations for different arguments. This concept allows programmers to use
notation that is closer to the domain in which the code is being employed,
making it more expressive and readable.}: \lstinline{X1 + X2 + ...}.

Similarly, the \lstinline{minus} function implements subtraction, which can
also be executed using the \lstinline{-} operator, as in \lstinline{X1 - X2 - ...}. 
A complete list can be accessed in the
\href{https://docs.octave.org/latest/Arithmetic-Ops.html}{Octave documentation}
\cite{octave}.

The excerpt presented in \Cref{lst-plus} shows how the \lstinline{plus}
function is used to sum vectors, executing significantly faster than its
counterpart that uses a loop to sum the vectors element by element. In the test
under investigation, it was observed that using the integrated function it was $470\times$
faster.

\begin{lstlisting}[language=octave, escapechar=|, label=lst-plus, caption={Example of executing the \lstinline{plus} function to perform the addition of two vectors. The function was used directly (line \ref{l-plus}), through operator overloading (line \ref{l-plus2}), and with a loop to perform element-by-element addition (line \ref{l-plus-loop}).}]
|\myOctave|X = [1:1E8];
|\myOctave\label{l-plus}|tic; plus (X, X); toc;
Elapsed time is 0.585023 seconds.
|\myOctave\label{l-plus2}|tic; X + X; toc;
Elapsed time is 0.587963 seconds.
|\myOctave\label{l-plus-loop}|tic; for i = 1:length (X), X(i) + X(i); endfor; toc;
Elapsed time is 274.972 seconds.
\end{lstlisting}

In addition to the performance gain during execution, bear in mind how the notation
using the operator \lstinline{+} is more intuitive and closer to mathematical
notation. This abstraction eliminates the underlying details of how vector
addition is implemented internally. This process is essential for handling
complex problems, as an abundance of details would obscure our true
understanding of the problem.
 
\section{Notation as a tool for introspection}\label{sec-notation}

\epigraph{Die Grenzen meiner Sprache bedeuten die Grenzen meiner Welt}{Ludwig
Wittgenstein, \emph{Tractatus Logico-Philosophicus}}

Iverson \cite{iverson1980} proposes that the notation used in programming
should also serve as a tool for introspection, easing the understanding
and analysis of algorithms and becoming an appropriate way to express the
addressed problem. He focused on developing a notation that was concise,
expressive, and close to the mathematical notation used in scientific and
mathematical problems. In this spirit, Iverson developed the APL (A Programming
Language), from which GNU Octave can be considered a descendant. Octave, being
a programming language focused on scientific and numerical calculations, aligns
with this proposal by utilizing notation similar to the mathematics used in
linear algebra. As seen earlier in \Cref{lst-plus}, Octave allows matrix and
vector operations to be performed in a concise and efficient manner. We will
now explore further parallels between this language and mathematical notation.

The definition of a vector in Octave is quite similar to its algebraic
representation. For example, the mathematical notation to define a vector can
be expressed as:
$X = \big[\begin{smallmatrix}
1 & 2 & 3 & 4 & 5 & 6 & 7 & 8 & 9
\end{smallmatrix}\big]$.
Similarly, in Octave, the same vector would be defined by: 
\lstinline{X = [1 2 3 4 5 6 7 8 9]}.

To calculate the sum of the vector elements, the mathematical
notation would be: $S = \sum_{i=1}^{|X|} x_i$, where $|X|$ represents the
number of elements in the vector $X$ and $x_i$ is the $i$-th element of the
vector. In Octave, the same operation can be performed with the built-in
function: \lstinline{S = sum (X)}. This function implicitly performs a loop
summing the elements of \lstinline{X}, but is implemented in \CC{} and
compiled, making it significantly more efficient than manually writing a loop
in an Octave script. Many other basic functions also follows this approach
to leverage their performance.
As a general rule, whenever possible, it is recommended
to use these built-in functions.

The previous representation is noteworthy for being more intuitive than using a loop:
\lstinline{S = 0; for i = 1 (X), S += X(i); endfor}. In this example, we used
some syntactic expressions that are not available in MATLAB, such as the
\lstinline{+=} operator and the use of the \lstinline{endfor} statement to
close the scope of the \lstinline{for} loop.
GNU Octave also introduced other small differences in its syntax compared to
MATLAB. A brief description of these differences can be found in the
\href{https://wiki.octave.org/wiki/index.php?title=Differences_between_Octave_and_Matlab}{GNU
Octave Wiki} \cite{octavewiki}.

The solution to a system of linear equations can be mathematically represented
as: $Ax=b$, where $A$ is the coefficient matrix, $x$ is the vector of unknowns,
and $b$ is the vector of known constants. Octave uses the \lstinline{\}
operator, known as the left division operator, to express the solution of
linear systems concisely and efficiently. By applying this operator, we can
solve the linear system and thus find the values of the unknowns: 
\lstinline{x = A \ b}. The \lstinline{\} operator is equivalent to the function
\lstinline{mldivide}. A complete list of overloaded functions and operators can
be found in the
\href{https://docs.octave.org/latest/Operator-Overloading.html}{Octave
documentation} \cite{octave}.

Hundreds of years of development have shaped the mathematical notation used
today \cite{cajori2003}. This notation has played a crucial role in scientific
advancements throughout history, serving as a lingua franca for the
communication and expression of complex mathematical concepts. Similarly, the
computational tools available may adopt distinct notations and are also
constantly evolving. Each problem and its solution can be more effectively
expressed through programming languages that adopt paradigms suited to
represent them. As we discussed earlier, GNU Octave, for example, was designed
to be a convenient tool for solving linear algebra problems and handling
matrices, making the interaction with the world of mathematics and computing
more effective and intuitive.

In the following sections, we will present some aspects of vector notation in
Octave. Through examples, we will illustrate how using this notation closely
resembles mathematical representation, highlight the performance gains achieved
by adopting this notation, for which Octave is optimized, and demonstrate how
its use makes our code less prone to errors.

  \subsection{Linear Algebra}\label{sec-algebralinear}

Linear Algebra plays a fundamental role in the sciences, serving as a universal
language to describe and solve a wide range of problems in fields such as
engineering, physics, computer science, statistics, and others
\cite{strang2005}.

Octave was created to address complex Linear Algebra problems efficiently and
intuitively. One of the main contributing factors is its use of a notation that
closely resembles the mathematical notation used in Linear Algebra. Thus,
Octave serves as a tool for introspection on the mathematical representation of
problems, enabling the direct representation of concepts in its language
without the need for complex conversions between mathematical notation and its
computational implementation.

To illustrate some aspects, consider two vectors in $\mathbb{R}^n$:
$\mathbf{a} = [a_1, a_2, \cdots, a_n]^\intercal$ and $\mathbf{b} = [b_1, b_2, \cdots, b_n]^\intercal$.
The dot product between them is given by
\begin{equation*} 
\mathbf a \cdot \mathbf b = \mathbf a^\intercal \mathbf b = \sum_{i=1}^n a_i b_i = a_1 b_1 + a_2 b_2 + \cdots + a_n b_n. 
\end{equation*}

In \Cref{lst-prod-escalar}, we present the calculation of this dot product. In
line \ref{ln-erro}, we made the mistake of not defining the variable
\lstinline{r}, where we accumulate the sum of the terms $a_i b_i$ for each $i$
from $1$ to $n$. The notation adopted in line \ref{ln-prodsc} reflects the
mathematical notation $\mathbf a^\intercal \mathbf b$, making it more compact
and thus less prone to implementation errors. Also, note (see lines
\ref{ln-prdtm} and \ref{ln-prdsctm}) how execution using vector notation is
approximately 700 times faster.

\begin{lstlisting}[language=octave, escapechar=|, label=lst-prod-escalar, caption={Scalar product between two vectors.}]
|\myOctave|a = [1 2 3]'; b = [3 2 1]';
|\myOctave\label{ln-erro}|for i = 1:length (a), r += a(i)*b(i); endfor
error: in computed assignment A OP= X, A must be defined first
|\myOctave|r = 0; for i = 1:length (a), r += a(i)*b(i); endfor
|\myOctave\label{ln-prodsc}|r = a'*b;
|\myOctave\label{ln-defarand}|a = rand (1E6,1); b = rand (1E6,1);
|\myOctave\label{ln-prdtm}|tic; r = 0; for i = 1:length (a), r += a(i)*b(i); endfor; toc;
Elapsed time is 2.76726 seconds.
|\myOctave\label{ln-prdsctm}|tic; r = a'*b; toc;
Elapsed time is 0.00390911 seconds.
\end{lstlisting}

In \Cref{sec-pca}, we will present principal component analysis as a backdrop
to once again address vector notation, show its proximity to mathematical notation,
and how such an approach provides greater intuition about the problem.

   \subsubsection{Principal Component Analysis}\label{sec-pca}

To illustrate how mathematical vector notation resembles the notation used in
Octave, we will present Principal Component Analysis (PCA) as an example. PCA
is an orthogonal linear transformation that allows data to be projected into a
new coordinate system, where most of the variance in the data is represented in
a few coordinates. This means that the first dimension will contain the most
variance and will be called the principal component, while the second dimension
will contain the second most variance, and so on.

To perform this transformation, we need to find the matrix of eigenvectors of
the correlation matrix of the data. Let $\mathbf{X}$ be the $m \times n$
matrix, with the data samples in  $\mathbb{R}^n$ arranged along the rows of
$\mathbf{X}$; then the autocorrelation matrix of $\mathbf{X}$ is given by
\begin{equation*}\label{eq-covmtx}
 \mathbf{S}_{\mathbf{X}} = \frac{\mathbf{X} \mathbf{X}^\intercal}{n-1}.
\end{equation*}
The orthogonal matrix  $\mathbf{P}$ that performs the PCA transformation is the
matrix where the eigenvectors of $\mathbf{S}_{\mathbf{X}}$ are arranged as the
rows of $\mathbf{P}$. By applying the transformation, or projecting the data
onto the rows of $\mathbf{P}$, we obtain the matrix $\mathbf{Y}$:
\begin{equation*}\label{eq-pca}
 \mathbf{Y} = \mathbf{P}^\intercal \mathbf{X}.
\end{equation*}
The translation to Octave is straightforward, as illustrated in \Cref{lst-pca}.
This highlights the similarity between the mathematical notation and the vector
notation used in Octave, thus facilitating introspection on the transformation
through the analysis of its implementation in this language.

An important preprocessing step before applying PCA is to subtract the mean
across each dimension of the data. This step is crucial to eliminate any bias
or offset that may exist in the data. It ensures that the PCA directions
accurately represent the variation structure within the data, without being
affected by offsets in the data. This step is accomplished through the code
snippet \lstinline{X -= mean (X, 2)} in \Cref{lst-pca}. This implementation
makes this normalization step concise, intuitive, and efficient.

For greater conciseness and efficiency, this notation also employs operator
broadcasting, a concept that will be discussed in \Cref{sec-broadcast}.
Calculating the mean for each dimension and normalizing the mean through a loop
would be more complex, this would be implemented as follows: 
\lstinline{mm = mean (X, 2); for i = 1 (mm), data(i,:) -= mm(i); endfor}.

\begin{lstlisting}[language=octave, escapechar=|, label=lst-pca, caption={Principal Component Analysis. Function simplified definition and example of usage.}]
|\myOctave|function Y = pca (X),
> n = columns (X);
> X -= mean (X, 2);
> S = X * X' / (n-1);
> [P, _] = eig(S);
> Y = P'*X;
> endfunction
|\myOctave|theta = pi/4;
|\myOctave|R = [cos(theta) -sin(theta); sin(theta)  cos(theta)];
|\myOctave|X = R*(randn (2, 100) .* [1; 0.1]);
|\myOctave|Y = pca (X);
|\myOctave|figure; subplot (1,2,1); plot (X(1,:), X(2,:), '.'); subplot (1,2,2); plot (Y(1,:), Y(2,:), '.');
\end{lstlisting}  \subsection{Indexing}\label{sec-idx}

Index expressions allow referencing elements in a matrix using indices. The
indices can be scalars, vectors, or ranges. We can also use the special
operator `\lstinline{:}' to express a complete row/column or to describe a
range. For example, \lstinline{2:8} describes the range from two to eight
($[2,8]$). We can also use the keyword `\lstinline{end}' to denote the last
element, thus eliminating the need to call the functions \lstinline{size} or
\lstinline{length}.

An index expression consists of a pair of parentheses \lstinline{( ... )}
enclosing $n$ expressions separated by commas, where each of these represents
an index associated with each of the $n$ dimensions of the object. It is also
possible to use a linear index to list the elements of a matrix. In this case,
we must remember that the elements are organized first along the first index
(rows), then along the second (columns), and so on.

To exemplify the use of index expressions, consider the two-dimensional magic
matrix\footnote{A magic matrix, or magic square, is the name given to the
square matrix in which the sum of the rows, columns, or main diagonals are
equal. Note that the matrix in the given example has a result of 34 for any of
the sums.} \lstinline{M} of size $4 \times 4$ defined in
\Cref{lst-magic-square}.

\begin{lstlisting}[language=octave, escapechar=|, label=lst-magic-square, caption={Magic matrix used in examples.}]
|\myOctave|M = magic (4);
|\myOctave|disp (M)
   16    2    3   13
    5   11   10    8
    9    7    6   12
    4   14   15    1
\end{lstlisting}

We can index the elements of a matrix using a linear index, as exemplified in
lines \ref{ex-ind1}, \ref{ex-ind2}, and \ref{ex-ind3}, or using two indices
(row and column), as illustrated in lines \ref{ex-ind4}, \ref{ex-ind5}, and
\ref{ex-ind6}.

\begin{multicols}{2}[\captionof{lstlisting}{Indexing expression usage examples.}]
\begin{lstlisting}[language=octave, escapechar=|, label=lst-exp-index]
|\myOctave\label{ex-ind1}|disp (M(3:7))
    9    4    2   11    7
|\myOctave\label{ex-ind2}|disp (M(12:end))
   15   13    8   12    1
|\myOctave\label{ex-ind3}|disp (M([1,3,5]))
   16    9    2
|\columnbreak|
|\myOctave\label{ex-ind4}|disp (M(1:2,2:3))
    2    3
   11   10
|\myOctave\label{ex-ind5}|disp (M(end,end-1:end)
   15    1
|\myOctave\label{ex-ind6}|disp (M(3,:))
    9    7    6   12
\end{lstlisting}\end{multicols}

Using index expressions makes data manipulation easier, allows you to use
vector operations on matrix elements, and also makes the code more readable,
efficient and concise. Learn more about
\href{https://docs.octave.org/latest/Index-Expressions.html}{index expressions}
in the Octave documentation.

  \subsubsection{Logic indexing}\label{sec-idx-logic}

Another very useful form of indexing is logical indexing. In this type of
indexing, we use a logical vector to extract the non-zero values corresponding
to this logical vector.

We will again use the magic matrix defined in
\Cref{lst-magic-square} as example. Suppose we want to extract the elements of
\lstinline{M} that are less than $8$. In line \ref{l-idxM}, we can see that the
expression \lstinline{M < 8} returns $1$ (one) for the elements that satisfy
this condition and $0$ (zero) for the others. We can use this logical index
vector to extract the desired elements, as shown in line \ref{l-idxM2}.

\begin{multicols}{2}[\captionof{lstlisting}{Logic indexing usage example.}]
\begin{lstlisting}[language=octave, escapechar=|, label=lst-logic-index]
|\myOctave\label{l-idxM}|disp (M < 8)
  0  1  1  0
  1  0  0  0
  0  1  1  0
  1  0  0  1
|\columnbreak|
|\myOctave\label{l-idxM2}|disp (M(M < 8)')
   5   4   2   7   3   6   1
\end{lstlisting}\end{multicols}

Several functions in Octave, whose names begin with \lstinline{is}, return a
logical vector. For example, suppose we want to replace the \lstinline{NaN}
elements of a matrix with $0$ (zero). Using the magic matrix \lstinline{M} as
an example, in \Cref{lst-logic-index-nan}, we will introduce a \lstinline{NaN}
in its first element and then replace it with $0$ (zero). We use the function
\lstinline{isnan}, which returns a logical index matrix with null entries for
the elements that are not \lstinline{NaN} and $1$ (one) for the \lstinline{NaN}
elements.

\begin{multicols}{2}[\captionof{lstlisting}{Using \lstinline{isnan} function to remove \lstinline{NaN} elements.}]
\begin{lstlisting}[language=octave, escapechar=|, label=lst-logic-index-nan]
|\myOctave|M(1,1) = 0/0;
|\myOctave|disp (isnan (M))
  1  0  0  0
  0  0  0  0
  0  0  0  0
  0  0  0  0
|\myOctave|M(isnan (M))=0;
|\myOctave|disp (M)
    0    2    3   13
    5   11   10    8
    9    7    6   12
    4   14   15    1
\end{lstlisting}\end{multicols}

Using logical indexing, in addition to being practical, is more efficient. To
demonstrate, in \Cref{lst-logic-index2}, we will create a random vector with
integers between $1$ and $100$ to calculate the average of those with a value
greater than $50$. Without using logical indexing, we must loop through the
elements of the vector, check if each element is greater than $50$, sum them,
count how many elements we found, and then divide the sum by the number of
elements (see line \ref{l-mean50loop}). In line \ref{l-mean50}, we obtain the
same result, but using logical indexing. Note the simplicity and efficiency.
For a vector with one million numbers, the code using logical indexing was
nearly $200\times$ faster.

\begin{lstlisting}[language=octave, escapechar=|, label=lst-logic-index2, caption={Execution time test to demonstrate the efficiency of using logical indexing.}]
|\myOctave|r = randi(100, 1, 1E6);
|\myOctave\label{l-mean50loop}|tic; s = 0; c = 0; for i=1:length (r), if(r(i)>50) s+=r(i); c++; endif; endfor; m = s/c; toc;
Elapsed time is 2.29225 seconds.
|\myOctave\label{l-mean50}|tic; mean ( r(r > 50) ); toc;
Elapsed time is 0.0117271 seconds.
\end{lstlisting}

Two useful functions to use with logical indexing are the functions \lstinline{any} and
\lstinline{all}. They are used to determine whether any or all, respectively,
elements of a matrix satisfy a certain condition. For example, in \Cref{lst-logic-index3},
we use these functions to generate warnings if any element of a vector is less
than $10$ or if all elements are greater than $40$.

\begin{lstlisting}[language=octave, escapechar=|, label=lst-logic-index3, caption={Examples of using the functions \lstinline{any} and \lstinline{all}.}]
|\myOctave|if any (r < 10), warning ("There is a value less than 10!"); endif
|\myOctave|if all (r > 40), warning ("All values are greater than 40!"); endif
warning: All values are greater than 40!
\end{lstlisting}

   \subsubsection{Scanning the elements of a matrix}\label{sec-mscan}

In order to illustrate how vector notation and the use of predefined functions
facilitate understanding, problem-solving, and result in more efficient code,
we will show how these concepts can be applied to the problem of scanning the
elements of a matrix.

\begin{figure}[htbp]
\centering
\begin{minipage}[t]{0.25\textwidth}
\centering
\includegraphics[width=0.75\textwidth]{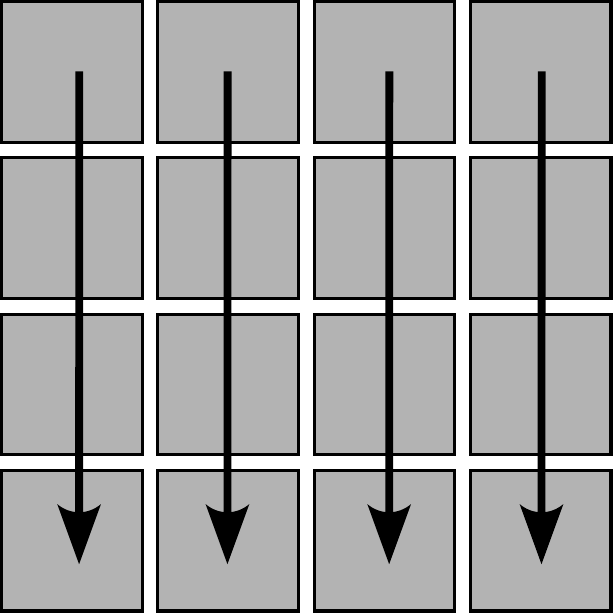}
\subcaption{Progressive scan.}\label{img-linearscan}
\end{minipage}
\hfill
\begin{minipage}[t]{0.25\textwidth}
\centering
\includegraphics[width=0.75\textwidth]{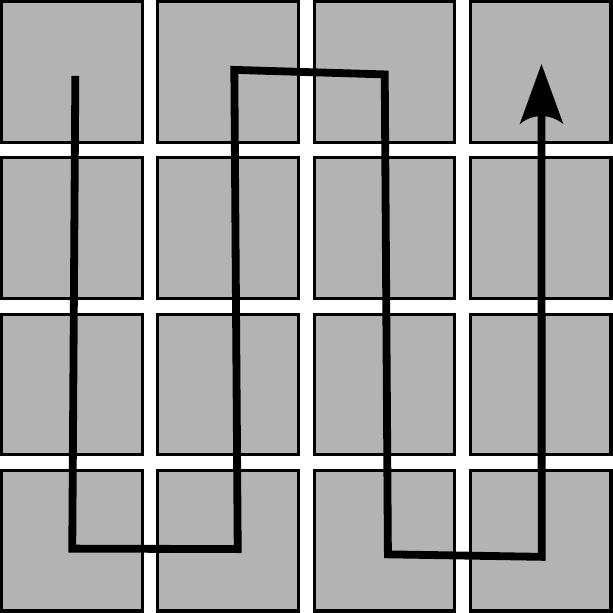}
\subcaption{Boustrophedon scan.}\label{img-boustrophedon}
\end{minipage}
\hfill
\begin{minipage}[t]{0.25\textwidth}
\centering
\includegraphics[width=0.75\textwidth]{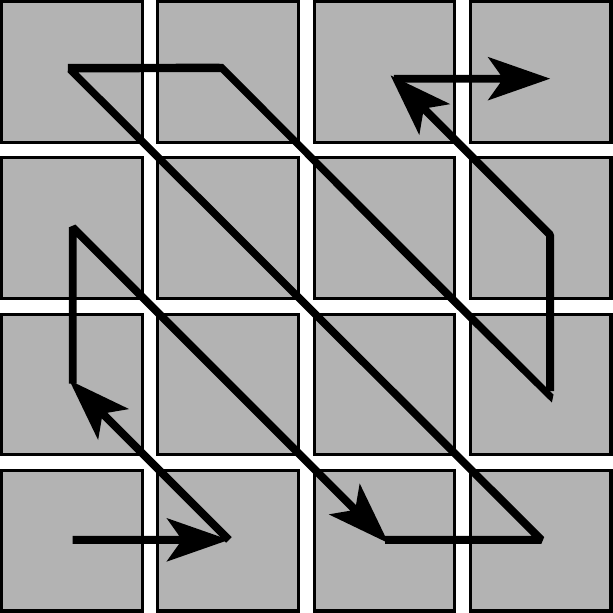}
\subcaption{Zig-zag scan.}\label{img-zigzag}
\end{minipage}
\caption{Three usual scanning techniques (directions were changed for simplicity).}
\label{fig-imgscan}
\end{figure}

\Cref{fig-imgscan} presents three types of matrix scanning that are commonly
used, especially in the context of digital image processing. The linear scan
(\cref{img-linearscan}) is the simplest, where rows/columns are read
sequentially. The boustrophedon scan (\cref{img-boustrophedon}) is similar to the
linear scan, but to avoid discontinuity when moving from one row/column to the
next, the reading direction is reversed. Finally, the zigzag scan
(\cref{img-zigzag}) aims to order the elements of the matrix following a zigzag
pattern. This last type of scan is used, for example, in the JPEG standard to
order the coefficients of the cosine transform. We adopt column reading as the
standard form, thus following the pattern adopted by Octave. For the zigzag
reading, we choose the starting point as the lower-left element and begin the
zigzag moving to the right. This choice was made solely to simplify the example
illustrated here.

To exemplify the scanning process, we will use the $4 \times 4$ magic matrix
given in \Cref{lst-magic-square} as an example. In a structured programming
language, to read the values of a matrix following the linear order
(\cref{img-linearscan}), we would use nested \emph{loops} to traverse the
elements of the matrix along its rows and columns. \Cref{lst-scan-loop-linear}
represents how this strategy can be implemented using Octave.

\begin{lstlisting}[language=octave, escapechar=|, label=lst-scan-loop-linear, caption={Loop to linearly scan a matrix.}]
|\myOctave|for j = 1:size (M,2), for i = 1:size (M,1), m((j-1)*size (M,2)+i) = M(i,j); endfor; endfor; disp (m)
   16    5    9    4    2   11    7   14    3   10    6   15   13    8   12    1
\end{lstlisting}

However, we can use index expressions (see \Cref{sec-idx}) to simplify the
linear scanning of the matrix elements. See the implementation presented in
\Cref{lst-scan-linear}, where the simple expression \lstinline{M(:)} results in
a column vector. This expression is a `code idiom'\footnote{A `code idiom' or
'programming idiom' is a fragment of code that performs a specific semantic
role and is frequently repeated in various codes.} found very often; it is
faster and generally clearer than performing a loop or calling the
\lstinline{reshape} function.

\begin{lstlisting}[language=octave, escapechar=|, label=lst-scan-linear, caption={Linear scanning a matrix using a simple index expression.}]
|\myOctave|disp (M(:)')
   16    5    9    4    2   11    7   14    3   10    6   15   13    8   12    1
\end{lstlisting}

In \Cref{lst-scan-loop-prog}, we present a structured example for implementing
the boustrophedon scanning of the matrix elements. Essentially, it is similar to
the linear scan, but we must reverse the reading order in the even columns.

\begin{lstlisting}[language=octave, escapechar=|, label=lst-scan-loop-prog, caption={Boustrophedon scannig a matrix using nested \emph{loops}.}]
|\myOctave|for j = 1:size (M,2), for i = 1:size (M,1),
> if (mod (j,2) == 1), v = M(i,j);
> else, v = M(size (M,1)-i+1,j);
> endif;
> m((j-1)*size (M,2)+i) = v;
> endfor; endfor;
|\myOctave|disp (m);
   16    5    9    4   14    7   11    2    3   10    6   15    1   12    8   13
\end{lstlisting}

This scanning process can be expressed more simply by using index expressions
to select the even columns and the function \lstinline{flipud} to reverse the
order of the elements in those columns. By observing the \emph{help} and the
code of this function, we find that it simply generates a reverse index vector
(\lstinline{size (x,dim):-1:1}) to index the elements. Thus, lines
\ref{line:flipup} and \ref{line:invindex} in \Cref{lst-scan-prog} are
equivalent; however, the first uses an extra level of abstraction, making the
code more intuitive and avoiding greater complexities in dealing with the
indices that reference the elements of the even columns.

\begin{lstlisting}[language=octave, escapechar=|, label=lst-scan-prog, caption={Boustrophedon scanning a matrix using index expression (only or also using the \texttt{flipud} function).}]
|\myOctave\label{line:flipup}|M(:,2:2:end) = flipud (M(:,2:2:end)); disp (M(:)')
   16    5    9    4   14    7   11    2    3   10    6   15    1   12    8   13
|\myOctave\label{line:invindex}|M(:,2:2:end) = M([size (M,2):-1:1],2:2:end); disp (M(:)')
   16    5    9    4   14    7   11    2    3   10    6   15    1   12    8   13
\end{lstlisting}

The last proposed example is the zigzag scan. This scanning method is more
complex to implement. We will start from the leftmost element at the bottom of
the matrix and proceed as illustrated in \Cref{img-zigzag}. In the first
implementation, presented in \Cref{lst-scan-zigzag}, we use a variable
\lstinline{c} that defines the direction of the scan; when it is negative, the
scan goes up a diagonal, and when it is positive, it goes down a diagonal. Upon
reaching the limits of the matrix, it is necessary to adjust the position
counters to continue traversing its elements.

\begin{lstlisting}[language=octave, escapechar=|, label=lst-scan-zigzag, caption={Zigzag scanning a matrix using loop and selection structure.}]
|\myOctave|function r = inrange(i,m),
> if m(1) <= i && i <= m(2), r = true;
> else, r = false;
> endif;
> endfunction
octave:16> function m = zigzag_loop(M),
> k = 1; c = 1; i = size (M,1); j = 1;
> while (inrange (i, [1 size(M,1)]) && (inrange (j, [1 size(M,2)]))),
>   m(k) = M(i,j); i = i+c; j = j+c; k++;
>   if i < 1 && j < 1, i++; j += 2; c *= (-1);
>   elseif i > size (M,1) && j > size (M,2), i -= 2; j--; c *= (-1);
>   elseif i < 1, i++; j += 2; c *= (-1);
>   elseif i > size (M,1), i--; c *= (-1);
>   elseif j < 1, j++; c *= (-1);
>   elseif j > size (M,2), j--; i -= 2; c *= (-1);
>   endif;
> endwhile
> endfunction
\end{lstlisting}

The function for zigzag scanning of the matrix can be simplified by using the
function \lstinline{spdiags} and the function \lstinline{flipud} (see
\Cref{lst-zigzag}). The former is used to extract the diagonals of the matrix,
while the latter is used to alternately reverse the direction of reading,
either going up or down along the diagonals. This implementation is more
concise and comprehensible due to the higher level of abstraction employed.

\begin{lstlisting}[language=octave, escapechar=|, label=lst-zigzag, caption={Zigzag scanning a matrix using \texttt{spdiags} and \texttt{flipud} functions.}]
|\myOctave|function m = zigzag(M),
> ind = reshape (1:numel(M), size (M));
> ind = spdiags (ind);
> ind(:,2:2:end) = flipud (ind(:,2:2:end));
> ind(ind==0) = [];
> m = M(ind)';
> endfunction
|\myOctave|disp (zigzag (M))
   32   24   35   19   32   37    4   30   31   48    9   32   39   40   46   35
\end{lstlisting}

Finally, we will compare the execution time of the zigzag scanning for the two
proposed implementations. The result is presented in \Cref{lst-zigzag-time}.
Note how the second implementation is approximately 6 times faster.

\begin{lstlisting}[language=octave, escapechar=|, label=lst-zigzag-time, caption={Comparison of the execution time of the two approaches for zigzag scanning the matrix.}]
|\myOctave|tic; for i=1:1E3, m=zigzag_loop (M); end; toc;
Elapsed time is 2.39824 seconds.
|\myOctave|tic; for i=1:1E3, m=zigzag (M); end; toc;
Elapsed time is 0.400645 seconds.
\end{lstlisting}

  \subsection{Broadcast}\label{sec-broadcast}

Broadcasting is a concept that refers to the ability to perform operations
between matrices of different sizes in a coherent and intuitive manner. In
simple terms, we say that the smaller array is broadcasted across the larger
array, applying a given operator along the way until they have compatible
dimensions. For example, if we have a matrix $X_{3 \times 3}$ and a vector
$Y_{1 \times 3}$, when we do \lstinline{X + Y}, the vector \lstinline{Y} will
be `copied'\footnote{The internal implementation of broadcasting reuses the
elements of the lower-dimensional matrix without the need to create copies of
its elements.} across the other dimension so that \lstinline{X + Y} can be
computed. That is, when one of the dimensions is equal to 1, the matrix with
that singleton dimension is replicated along the other dimension until it
matches the dimension of the matrix with which the operation is being
performed.

\begin{lstlisting}[language=octave, escapechar=|, label=lst-broadcast, caption={Applying \emph{broadcast} to the addition operator.}]
|\myOctave|X = [1 2 3; 4 5 6; 7 8 9];
|\myOctave|Y = [10 20 30];
|\myOctave|disp (x + y)
   11   22   33
   14   25   36
   17   28   39
\end{lstlisting}

Simpler examples of broadcasting are the operations performed with scalars.
When we perform the addition \lstinline{X + 10}, the comparison \lstinline{X > 5}, 
or \lstinline{min (X, 5)}, we are using broadcasting with scalars.

An example of an application with higher dimensionality is the conversion of an
RGB image to grayscale. An RGB image is a three-dimensional matrix where its
dimensions are rows, columns, and color components. In the case of an RGB
image, the third dimension has a length of 3. A grayscale image can be
represented by a two-dimensional matrix, where usually the value of each pixel
represents a shade of gray or, in this case, the luminance value associated
with the pixel. We will perform the conversion from color to grayscale using
the relative luminance contribution provided by each of the RGB components, the
primary colors red, green, and blue. The function \lstinline{rgb2gray} uses the
conversion as specified in Rec. ITU-R BT.601-7 \cite{rbt601-7}. According to
this recommendation, the luminance $Y'$ is given by
\begin{equation}\label{eq-rgb2gray}
    Y' =  0.299 R' + 0.587 G' + 0.114 B',
\end{equation}
where $Y'$ represents the luminance with gamma correction\footnote{Gamma
correction is usually used to account for the non-linear response of human
vision.} and the variables $R'$, $G'$, and $B'$ represent the color component
values also with gamma correction. To simplify, we will ignore gamma correction
and see how we can perform such a calculation using broadcasting. The
implementation of this example appears in \Cref{lst-img-gray}.

\begin{lstlisting}[language=octave, escapechar=|, label=lst-img-gray, caption={Using \emph{broadcast} to convert a color image into grayscale.}]
|\myOctave|url = "https://sipi.usc.edu/database/download.php?vol=misc&img=4.1.05";
|\myOctave|urlwrite (url, "house.tiff");
|\myOctave|img = imread ("house.tiff");
|\myOctave|img_gray = sum(img .* permute ([0.299, 0.587, 0.114], [1, 3, 2]), 3);
|\myOctave|figure; subplot (1,2,1); imshow (img); title ('original'); subplot (1,2,2); imshow (uint8 (img_gray)); title ('grayscale');
\end{lstlisting}

In the given example, we used the function \lstinline{permute} to permute the
dimensions of the matrix with the weight coefficients given in
\Cref{eq-rgb2gray}, that is, we applied a transpose\footnote{The transposition
    of a matrix is a simple permutation of the first and second dimensions,
    that is, \lstinline{permute (A, [2 1])}. It is therefore equivalent to applying
    the transposition operator to the matrix \lstinline{A}, hence equivalent to
    \lstinline{A'}.} 
generalization for matrices with more than two dimensions.

   \subsubsection{Euclidean Distance}

In this section, we present three implementations of functions to compute the
Euclidean distance between all points in a set. The points are represented by
the rows of the matrix \lstinline{P}.

\Cref{lst-distance1} shows the implementation using the structured programming
paradigm. In this implementation, we perform two \emph{loops} to select pairs
of points (rows of the matrix) and then another \emph{loop} to compute the
distance between these points. The second loop for selecting the
points starts at \lstinline{j = i}, taking advantage of the fact that distances
have the property of symmetry, something we could not apply if we were
computing a divergence, for example, the Kullback-Leibler divergence.

\begin{lstlisting}[language=octave, escapechar=|, label=lst-distance1, caption={Implementation using three loops to compute distance between pairs of points.}]
|\myOctave|function d = distance1 (P),
>  [N, M] = size (P);
>  for i = 1:N, for j = i:N,
>    dist = 0;
>    for k = 1:M,
>      dist += (P(i,k) - P(j,k)).^2;
>    endfor;
>    d(i,j) = d(j,i) = sqrt (dist);
>  endfor; endfor
> endfunction
\end{lstlisting}

In the next implementation, presented in \Cref{lst-distance2}, we use
\emph{broadcasting} and indexing to compute the distance from one point to
all others. Similarly to what we did before, we will still use symmetry to
simplify and perform only half of the computations.

\begin{lstlisting}[language=octave, escapechar=|, label=lst-distance2, caption={Implementation using broadcast and idexing to compute all distances from one point to the others, using one loop to change the reference point.}]
|\myOctave|function d = distance2 (P),
>   N = size (P,1);
>   for i = 1:N,
>     d(i:N,i) = sqrt (sum ((P(i:N,:) - P(i,:)).^2,2));
>     d(i,i:N) = d(i:N,i)';
>   endfor
> endfunction
\end{lstlisting}

Finally, we present a vectorized implementation that also uses
\emph{broadcast} to compute all distances between all points, including
the same pairs of points in reverse order, without leveraging the symmetry of
the problem.

\begin{lstlisting}[language=octave, escapechar=|, label=lst-distance3, caption={Implementation using broadcast to compute all distances at once.}]
|\myOctave|function d = distance3 (P),
>   d = permute (sqrt (sum ((P - permute (P, [3, 2, 1])).^2, 2)), [1,3,2]);
> endfunction
\end{lstlisting}

In \Cref{lst-distance-time}, we compare the execution time of each approach to
compute one million distances between each pair in a set of one thousand
points. Remark that the last implementation was significantly superior to the
others, even without utilizing the problem's symmetry. The first implementation
proved to be the least efficient, taking $200 \times$ the execution time of the
best implementation. The second shows a factor of $30 \times$, still
significantly greater than the best implementation.

\begin{lstlisting}[language=octave, escapechar=|, label=lst-distance-time, caption={Comparison of the execution time of each of the functions to compute the distances in a set of one thousand points.}]
|\myOctave|tic; d = distance1 (P); toc;
Elapsed time is 11.0255 seconds.
|\myOctave|tic; d = distance2 (P); toc;
Elapsed time is 1.65677 seconds.
|\myOctave|tic; d = distance3 (P); toc;
Elapsed time is 0.0536449 seconds.
\end{lstlisting}

Broadcast allows for shorter, more intelligible, and easily maintainable code,
in addition to providing greater efficiency in terms of execution time and
memory.  For functions that don't naturally support \emph{broadcasting}, we can
use the function \lstinline{bsxfun} to coerce them to apply \emph{broadcast}.
Read more about the topic in the
\href{https://docs.octave.org/latest/Broadcasting.html}{Octave documentation}.

  \subsection{Function Handles}\label{sec-functionhandle}

A \emph{function handle} is an object that represents a function. This makes it
possible to assign functions to variables and pass them as arguments to other
functions. We can thus create anonymous functions\footnote{Anonymous functions
are functions that are not stored in a program file, but are stored in a
variable of type \texttt{function\_handle}. A simple example of defining an
anonymous function is given as follows: \lstinline{quad = @(x) x.^2;}, where
the variable \lstinline{quad} stores a handle for a function that implements
the square function.} for specific tasks, pass them as arguments to other
functions, or create higher-order functions, which are functions that operate
on other functions. In this way, we can treat functions as first-class
objects\footnote{In programming language design, a first-class object is an
entity that can appear in expressions, be assigned to variables, and be used as
an argument.}, allowing for greater flexibility and abstraction, enabling us to
create more generic and reusable code.

As an example, in \Cref{lst-rplnaneg}, we will create a conditional function to
replace \lstinline{NaN} or negative values with zero.

\begin{lstlisting}[language=octave, escapechar=$, label=lst-rplnaneg, caption={Example of an anonymous function to replace negative values and \lstinline{NaN} with zero.}]
$\myOctave$replace_negative = @(x) (x < 0) .* 0 + (x >= 0) .* x;
$\myOctave$disp (replace_negative ([-1 1 -2 2 -3 3]));
   0   1   0   2   0   3
$\myOctave$replace_neg_nan = @(x) ifelse(isnan(x) | x < 0, 0, x);
$\myOctave$disp (replace_neg_nan ([0 1 2 -1 NaN 3 -2 4]));
   0   1   2   0   0   3   0   4
\end{lstlisting}

\subsubsection{k Nearest Neighbors}\label{sec-knn}

In the example presented in \Cref{lst-nn}, we create a function that uses
another function, passed as an argument, to compute the distances between
points (for each element of \lstinline{Y}, the distances to all elements of
\lstinline{X} are computed). The function returns the index of the element in
\lstinline{X} that is closest to the corresponding element in \lstinline{Y}, as
well as the distance between them. The \lstinline{nearestneighbor} function
takes as one of its parameters the distance function, which should be
implemented as a handle to an anonymous function. In the example, we present
two common distances: Euclidean distance and Manhattan distance, given in
\Cref{ed-dist-eucl,eq-dist-manh}, respectively.

\begin{minipage}[m]{0.45\textwidth}
  \begin{equation*}\label{ed-dist-eucl}
      d_e (x,y) = \sqrt{ \sum_{i=1}^{n} (x_i - y_i)^2 },
  \end{equation*}
\end{minipage}
\hfill
\begin{minipage}[m]{0.45\textwidth}
  \begin{equation*}\label{eq-dist-manh}
      d_m (x,y) = \sum_{i=1}^{n} \vert x_i - y_i \vert .
  \end{equation*}
\end{minipage}

\begin{lstlisting}[language=octave, escapechar=|, label=lst-nn, caption={Example of a function to return the nearest neighbor, allowing the use of different distance metrics. In this example, we define two anonymous functions to implement the calculation of the Euclidean distance and the Manhattan distance.}]
|\myOctave|manhattan = @(X,Y) permute (sum (abs (X - permute (Y, [3 2 1])), 2), [1 3 2]);
|\myOctave|euclidean = @(X,Y) permute (sqrt (sum ((X - permute (Y, [3 2 1])).^2, 2)), [1 3 2]);
|\myOctave|function [idx, d] = nearestneighbor (X, Y, distance)
> > >    D = distance (X, Y);
>    [idx, d] = min (D, [], 2);
> endfunction
|\myOctave|X = [0 0; 1 1; 2 2]; Y = [0 1; 2 1];
|\myOctave|[idx, d] = nearestneighbor (X, Y, manhattan);
|\myOctave|[idx, d] = nearestneighbor (X, Y, euclidean);
\end{lstlisting}   \subsubsection{Discrete Cosine Transform in Image Compression}

Some Octave functions require a user-defined function as an argument. This is
the case for the \lstinline{blockproc} function (from the \emph{image}
package), presented in \Cref{lst-dct}. This function applies the provided
function to blocks of the matrix given as an argument. In this example, we will
apply the DCT\footnote{DCT stands for \emph{Discrete Cosine Transform}, an
    orthonormal transform commonly used in multimedia signal compression,
    thanks to its energy compacting characteristics and its ability to handle
    only real numbers when applied to real signals.} to $8\times 8$ blocks of a
matrix (this is one of the steps used in JPEG compression). To achieve this, we
will define the 2D DCT for $8\times 8$ matrices using an anonymous function
that will be passed to the \lstinline{blockproc} function. In line
\ref{l-dctmtx}, we define the DCT matrix using the \lstinline{dctmtx} function
(which belongs to the \emph{signal} package). Next, in line \ref{l-dct2d}, we
define the 2D DCT operation on a matrix (mathematically expressed as
$\mathbf{Y} = \mathbf{C} \mathbf{X} \mathbf{C}^\intercal$, where $\mathbf{X}$
is the data matrix, $\mathbf{C}$ is the DCT matrix, and $\mathbf{Y}$ is the
result of applying the DCT to the matrix $\mathbf{X}$). Next, we'll apply the
2D DCT defined in line \ref{l-blockproc} to image blocks using the
\lstinline{blockproc} function. Finally, visualization is done in line
\ref{l-vis-gray}.

\begin{lstlisting}[language=octave, escapechar=|, label=lst-dct, caption={Example of using an anonymous function to define the 2D DCT and utilize it in one of the steps of image compression.}]
|\myOctave|pkg load signal; pkg load image;
|\myOctave\label{l-blksz}|blk_sz = 8;
|\myOctave\label{l-dctmtx}|T = dctmtx (blk_sz);
|\myOctave\label{l-dct2d}|dct2d = @(x) T * x * T';
|\myOctave\label{l-blockproc}|img_dct = blockproc (img_gray, [blk_sz blk_sz], dct2d);
|\myOctave\label{l-vis-gray}|figure; colormap ('gray'); imagesc (abs (img_dct));
\end{lstlisting}

In the presented examples, the adopted notation contributes to the efficiency
and conciseness of the code. The code readability is improved, facilitating the
introspection process concerning the proposed problem. Along with improved code
efficiency, we will also benefit from enhanced reusability. In the example
from \Cref{lst-nn}, we see how straightforward it is to use different metrics.
Similarly, in \Cref{lst-dct}, we can easily swap the transform or adjust the
block sizes. This provides us with flexibility and structure for efficiently
analyze and compare various scenarios related to our study.

  \subsection{Other Common Functions in Code Vectorization}

Several Octave functions are particularly useful when adopting a vectorized
approach, enabling elegant and efficient execution of complex matrix
operations. \Cref{tbl-fnc-vec} lists key functions that help eliminate explicit loops, as
well as manipulate, transform, expand, or sort data, enhancing both the
efficiency and readability of your code. While some of these functions are
demonstrated in the examples throughout this work, others are left for the
reader to explore through the \emph{help} documentation and additional examples
available online.

\begin{table}[htpb]
\centering
\caption{List of Common Functions for Code Vectorization.}\label{tbl-fnc-vec}
\begin{tabularx}{\linewidth}{lX}
\toprule
function & description \\
\midrule
\lstinline|all| & \small{determines if all elements in the vector are non-zero} \\
\lstinline|any| & \small{determines if at least one element in the vector is non-zero} \\
\lstinline|cat|, \lstinline|horzcat|, \lstinline|vertcat| & \small{concatenate matrices} \\
\lstinline|deal| & \small{copy input parameters to corresponding output parameters} \\
\lstinline|diff| & \small{moving difference} \\
\lstinline|permute|, \lstinline|ipermute| & \small{permute the dimensions of a matrix} \\
\lstinline|prod| & \small{product} \\
\lstinline|repelems| & \small{replicate elements of a matrix} \\
\lstinline|repmat| & \small{replicate matrix} \\
\lstinline|reshape| & \small{rearrange elements of the matrix} \\
\lstinline|shift| & \small{circularly shift the elements of the matrix} \\
\lstinline|sort| & \small{sort the elements of the matrix} \\
\lstinline|sum|, \lstinline|cumsum|  & \small{sum and cumulative sum of the elements of a matrix} \\
\lstinline|sub2ind|, \lstinline|ind2sub| & \small{conversion between multidimensional index and linear index} \\
\lstinline|unique| & \small{returns the unique elements} \\
\bottomrule
\end{tabularx}
\end{table}

\section{Conclusion}\label{sec-conclusao}

\epigraph{I can't go back to yesterday because I was a different person then.}
{Lewis Carroll, \emph{Alice's Adventures in Wonderland}}

This article explores the relationship between notation and introspection,
using GNU Octave as a programming tool for problem-solving. Through Octave's
language, we highlight its focus on effectively representing complex
mathematical problems. Beyond notation, we examine how the language's features
and built-in functions contribute to more efficient and streamlined
programming. Mastery of vector notation can shift our perspective, offering a
clearer and more direct understanding of both problems and their solutions.

Octave's adoption of a notation closely aligned with linear algebra enables
programmers to express their intentions clearly and concisely. This approach
not only simplifies coding but also enhances code readability and supports
deeper introspection. With features like operator overloading, indexing,
multidimensional array handling, broadcasting, and function handles, Octave
provides a platform that naturally represents mathematical problems. This
flexibility allows for a more abstract and adaptable approach to
problem-solving, echoing Iverson's \cite{iverson1980} belief that notation
should serve as a tool to expand our cognitive abilities.

Ultimately, Octave's approach illustrates that notation is not merely a tool
for describing problems; it is also a means of engaging with them
intellectually. By examining the relationship between notation and
introspection, we aim to inspire programmers to address complex challenges with
greater effectiveness, clarity, and elegance. As programming continues to
evolve, notation remains a crucial pillar, empowering developers to translate
their visions into reality and transform problems into innovative solutions.

\bibliographystyle{plain} \bibliography{article}    

\end{document}


\maketitle

\begin{abstract}
    A notação vetorial adotada pelo GNU Octave desempenha um papel significativo
como uma ferramenta para introspecção, alinhando-se à visão de \textcite{iverson1980}.
Ele acreditava que, assim como a matemática, uma linguagem de
programação deve ser uma ferramenta de pensamento eficaz para representar e
raciocinar sobre problemas que desejamos abordar.  Este trabalho tem como
objetivo explorar o uso da notação vetorial no GNU Octave, por meio da análise
de operadores e funções, proporcionando uma maior proximidade com a notação
matemática e aprimorando a eficiência do código. Abordaremos conceitos
fundamentais, como indexação, \emph{broadcasting} e \emph{function handles}, e
apresentaremos casos de estudo para uma compreensão mais aprofundada desses
conceitos. Ao adotar a notação vetorial, o GNU Octave se torna uma poderosa
ferramenta para cientistas e engenheiros, permitindo que eles expressem e
resolvam problemas complexos de maneira mais eficaz e intuitiva.
   
    -----

    \selectlanguage{english}
    The vector notation adopted by GNU Octave plays a significant role as a tool
for introspection, aligning itself with the vision of \textcite{iverson1980}. He
believed that, just like mathematics, a programming language should be an
effective thinking tool for representing and reasoning about problems we wish
to address. This work aims to explore the use of vector notation in GNU Octave
through the analysis of operators and functions, providing a closer alignment
with mathematical notation and enhancing code efficiency. We will delve into
fundamental concepts such as indexing, broadcasting, and function handles, and
present case studies for a deeper understanding of these concepts. By adopting
vector notation, GNU Octave becomes a powerful tool for scientists and
engineers, enabling them to express and solve complex problems more effectively
and intuitively.
     \selectlanguage{portuguese}

\keywords{GNU Octave, Notação Vetorial, Eficiência de Código, Linguagem, Introspecção.}
\end{abstract}

\section{Introdução}\label{sec-intro} 
\epigraph{Uma das razões, ou melhor, uma explicação para o fato das linguas se
diferirem é que as comunidades de fala variam na forma como utilizam suas
línguas e nas funções comunicativas que as línguas desempenham.\footnotemark
}{Florian Coulmas, \emph{Language adaptation}}
\footnotetext{Texto original: \emph{One reason, or rather one explanation of why languages differ is that speech communities differ with respect to the ways they use their language(s) and the communicative functions that languages fulfill}.}

Uma linguagem de programação é um sistema formal utilizado para escrever
programas computacionais ou fornecer instruções a um computador
\cite{abelson1996}. Os programas visam resolver um problema, realizar uma
simulação, tarefa ou função. As aplicações podem variar amplamente, abrangendo
desde a edição de texto e a composição musical até o controle do lançamento e
da trajetória de um foguete.  Cada uma dessas distintas tarefas pode ser
abordada por meio de diversas representações. As linguagens de programação
utilizam-se da sintaxe e semântica para criar abstrações sobre a definição,
manipulação de dados e estruturas e controle de fluxo de execução. Determinadas
linguagens de programação são mais propícias a lidar com certos tipos de
problemas em comparação com outras, levando à uma notação e solução mais
simples, maior eficiência e aprimoramento da nossa capacidade de pensar,
comunicar e resolver tais questões de maneira mais eficaz \cite{mcconnell2004}.

As ferramentas computacionais podem ser utilizadas para complementar o pensamento
matemático de diversas formas. Primeiramente como uma forma de conceitualizar 
e construir o raciocínio, elaborar exemplos ou buscar contraexemplos, fornecer
dados e simulações que ajudariam a indicar conjecturas e até mesmo auxiliar nas
demonstrações ou provar teoremas\footnote{O teorema das quatro cores foi o primeiro
teorema provado utilizando computação. Kenneth Appel e Wolfgang Haken reduziram
o problema a um número finito (embora grande) de alternativas que foram testadas
computacionalmente. Os computadores passaram a ser utilizados em diversas outras 
provas matemáticas como no caso do último teorema de Fermat e da conjectura de Kepler,
dentre outros} \cite{dubinsky2002}. 
\begin{quote} 
    ``... geralmente é verdade que, sempre que uma pessoa constrói algo num
    computador, uma construção correspondente é feita na mente da pessoa. É
    possível orquestrar essa correspondência fornecendo tarefas de programação
    numa linguagem de programação apropriada, concebida de modo que as
    construções mentais resultantes sejam ideias poderosas que melhorem o
    conhecimento e a compreensão matemática''\footnote{Texto original: 
	``\emph{Indeed, it is in general true that whenever a person constructs something on a
	    computer, a corresponding construction is made in the person’s mind. It is
	    possible to orchestrate this correspondence by providing programming tasks
	    in an appropriate programming language designed so that the resulting mental
	    constructions are powerful ideas that enhance the student’s mathematical
    	    knowledge and understanding}''.} \cite{dubinsky2002}. 
\end{quote}

Assim como a programação oferece diferentes formas de representar e resolver
problemas, a introspecção -- o ato de olhar para dentro -- também depende de uma
representação clara de nossas ideias. A ideia de Matriz Espelho no Mural,
inspirada na famosa frase \emph{mirror, mirror on the wall}, remete ao conceito de
introspecção, onde o espelho serve como ferramenta de reflexão interna. Na
clássica história de Branca de Neve, o espelho responde às perguntas da rainha,
revelando verdades ocultas. Analogamente, cada abordagem de programação age
como um diferente `espelho' para o pensamento, refletindo de maneiras distintas
e criando abstrações diversas para a resolução de problemas complexos. A
escolha da notação adequada a cada problema permite a elaboração de soluções
mais elegantes e eficientes.

Em cursos de graduação, usualmente utiliza-se paradigmas procedurais como forma
introdutória ao universo da programação \cite{deraadt2004,mason2024}\footnote{No estudo
realizado por \textcite{deraadt2004}, os instrutores relatam utilizar, em cursos
introdutórios à programação, um dos seguintes paradigmas: procedural, orientado
a objetos e funcional. Na maioria dos casos, a abordagem adotada é a
procedural, mesmo que para tanto utilize-se uma linguagem de programação
orientada a objetos. \textcite{mason2024} também apontam uma observação semelhante.}. Os procedimentos são delineados por uma série de passos
computacionais que devem ser seguidos para realizar tarefas específicas. Essa
abordagem visa simplificar a compreensão do conceito de programação para
iniciantes. Os conceitos adotados são mais acessíveis, de fácil assimilação e
requerem menor nível de abstração por parte dos estudantes. A abordagem
estrutural desempenha um papel fundamental ao transmitir os princípios de
organização do código por meio de funções e procedimentos. As linguagens
procedurais também facilitam a assimilação, introduzindo os elementos básicos
da programação, como variáveis, tipos de dados, estruturas de controle de fluxo
e funções. A natureza sequencial inerente a essas linguagens se assemelha ao
pensamento sequencial frequentemente empregado na resoluções de problemas, o
que promove a intuição de raciocínio lógico na construção algorítmica. A
abordagem procedural também demanda uma menor dose de abstração. Aparentemente, a conjunção de
todos esses fatores as tornam um paradigma de programação mais acessível e
descomplicado, sendo portanto adotado majoritariamente nos cursos introdutórios 
à programação. Entretanto, alguns estudos mostram que o paradigma orientado a objetos
pode ser utilizado em cursos introdutórios promovendo resultados similares 
àqueles observados quando a abordagem procedural é adotada \cite{wiedenbeck1999,vilner2007fundamental}.

A interseção entre a linguagem de programação e a notação matemática deve ser
observada, sobretudo quando estamos lidando com problemas que são modelados e
solucionados matematicamente. Como expressou \textcite{iverson1980} em sua
obra, ``a maioria das linguagens de programação são claramente inferiores à
notação matemática e são pouco utilizadas como ferramentas de pensamento de
maneiras que seriam consideradas significativas''\footnote{Trecho original:
\emph{``most programming languages are decidedly inferior to mathematical
notation and are little used as tools of thought in ways that would be
considered significant.''}} \cite{iverson1980}. Essa afirmação provocativa nos
leva a explorar como as linguagens de programação tradicionais contrastam com a
eficácia da notação matemática como uma ferramenta de expressão e raciocínio.
Neste contexto, destaca-se a linguagem de programação APL\footnote{Nome dado em
referência ao livro \emph{`A Programming Language'}.}, desenvolvida por Iverson
na década de 1960, que introduziu uma abordagem que se alinhava com o conceito
de \emph{Notation as a Tool of Thought} (Notação como Ferramenta de
Pensamento). A APL utilizava matrizes multidimensionais, operadores e funções
que levavam ao desenvolvimento de códigos concisos. Além da vetorização, a APL
foi pioneira em fornecer um ambiente de programação interativo, permitindo a
concepção do código e análise de resultados concomitante \cite{mcleod}. Linguagens de
programação como a APL são voltadas à computação científica e engenharia,
utilizam o paradigma da programação vetorial. A estrutura básica de dados são
vetores, explorando o fato de que seus elementos constituintes são de um mesmo
tipo ou similares. O paradigma vetorial é utilizado com o intuito de
generalizar e aplicar operadores a escalares, vetores, matrizes ou vetores de
dimensões superiores. Essa abordagem fornece maior nível de abstração, fornecendo
uma notação mais concisa e eficiente para representar problemas matemáticos,
permitindo assim ao programador pensar e operar em estruturas de dados ao invés
de se preocupar em realizar \emph{loops} explícitos sobre os escalares que compõem
tais estruturas. A abordagem vetorial também é inerentemente propícia ao
paralelismo devido às operações realizadas elemento a elemento. Ademais, linguagens
modernas de programação vetorial podem tirar proveito de processadores
multicore e GPUs para uma computação mais rápida\footnote{A programação
    vetorial, que envolve operações em arrays ou matrizes de dados, pode se
    beneficiar significativamente do uso de instruções SIMD (\emph{Single
	Instruction, Multiple Data}) como aquelas encontradas nas famílias de
	instruções SSE (\emph{Streaming SIMD Extensions}) e do uso de GPUs
	(\emph{Graphics Processing Units}) através de linguagens como CUDA
	(\emph{Compute Unified Device Architecture}) da NVIDIA ou OpenCL
    (\emph{Open Computing Language}) que é um padrão aberto e multiplataforma
para programação paralela.} \cite{sawadsitang2012}.

Uma linguagem de programação, como sistema de notação, deve ser concisa e
eficiente, devendo priorizar a comunicação da informação essencial, minimizando
a inclusão de detalhes desnecessários. Para tanto, devemos utilizar uma notação
compacta e uma representação de mais alto nível. Sob essa perspectiva que
\textcite{iverson1980} argumenta em favor da utilização da linguagem como
ferramenta do pensamento. ``Um esquema fundamental para conseguir isso é a
introdução de regras gramaticais pelas quais frases e sentenças significativas
podem ser construídas pela combinação de elementos do
vocabulário''\footnote{Trecho original: \emph{A fundamental scheme for
	achieving this is the introduction of grammatical rules by which meaningful
	phrases and sentences can be constructed by combining elements of the
	vocabulary}.} \cite{iverson1980}.

As linguagens de programação fornecem a abstração necessária para o
desenvolvimento de sistemas complexos. Essa abstração se dá através da
supressão de detalhes, permitindo a reutilização de componentes e ferramentas
de forma a tornar tratável a complexidade que seria expressar qualquer
programa diretamente em linguagem de máquina. Uma linguagem de programação
deve aprimorar os seguintes aspectos essenciais dos programas: leitura,
escrita, execução e raciocínio \cite{orchard2011four}. Esses princípios se
alinham à visão de Iverson de que uma notação eficaz e concisa é essencial em
uma linguagem de programação. As características de uma linguagem de
programação também se conectam à hipótese Sapir-Whorf\footnote{Segundo a
hipótese Sapir-Whorf, a língua determina e molda a cognição e a compreensão de
mundo do falante.}, em que a linguagem de programação pode facilitar ou
dificultar a representação de certos conceitos e a elaboração do raciocínio 
sobre eles.
Na comparação abaixo, note como a expressão através de uma linguagem de
programação é mais intuitiva do que a representação matemática para uma 
função por partes:
\begin{center}
\begin{minipage}[m]{0.25\textwidth}
    \lstinline{if x > 0, then y = 1; else y = 0.}
\end{minipage}
\hspace{2cm}
\begin{minipage}[m]{0.3\textwidth}
    \begin{equation*}
    y = 
    \begin{cases}
	1 & , x > 0 \\
	0 & , \text{cc.}
    \end{cases}
    \end{equation*} 
\end{minipage}
\end{center}
A representação matemática utiliza a ordem inversa daquela utilizada em linguagem natural,
além da utilização de convenções, abreviações e acrônimos. No exemplo anterior, `cc' é o acrônimo
utilizado para \emph{caso contrário}. Muitas vezes a condicional `se' também é omitida,
como na primeira linha da definição por partes de $y$.

Neste artigo pretendemos apresentar o uso do GNU Octave, utilizando suas
características de uma linguagem de programação vetorial e evidenciando como
todo seu potencial pode ser atingido e as vantagens em se utilizar a notação
vetorial. Poderemos assim alcançar uma melhor estruturação, organização e
introspecção na elaboração de algoritmos, além de maximizar o potencial da
linguagem e desenvolver códigos significativamente mais eficientes. Através de
exemplos e comparações, pretendemos ainda instruir o processo de migração do
paradigma procedural para o paradigma vetorial.  A exposição e esclarecimento
sobre o paradigma vetorial é importante uma vez que usualmente os estudantes de
cursos de graduação chegam ao final do curso utilizando GNU Octave/MATLAB
apenas sob o paradigma procedural.

\section{GNU Octave}\label{sec-octave}

\epigraph{Só podemos ver uma curta distância à frente, mas podemos ver muito
que precisa ser feito.\footnotemark}{Alan Turing, \emph{Computing machinery and
intelligence}} \footnotetext{Texto original: \emph{We can only see a short
distance ahead, but we can see plenty there that needs to be done.}}

O GNU Octave é um ambiente de computação numérica que utiliza uma linguagem de
programação de alto nível interpretada. Sua principal ênfase reside nos
cálculos matemáticos, análise e visualização de dados, resolução numérica de
problemas lineares e não-lineares, bem como no desenvolvimento de algoritmos
\cite{octave}. Oferecendo uma alternativa gratuita e de código aberto ao
MATLAB, o GNU Octave é cuidadosamente projetado para ser compatível com este
último. Consequentemente, torna-se uma excelente opção para estudantes e
acadêmicos que buscam uma alternativa viável sem os encargos associados às
licenças de software.

O Octave foi inicialmente concebido por John W. Eaton em 1992 \cite{octave},
passando a fazer parte do \href{https://www.gnu.org/}{Projeto GNU}, uma
iniciativa de software livre e de código aberto que foi lançada por Richard
Stallman em 1983. Sendo um software livre, o GNU Octave recebe contribuições de
uma diversidade de programadores.  Assim como o MATLAB, o Octave conta com
diversos pacotes. Estes são mantidos e desenvolvidos pela comunidade, o que
pode resultar em uma variedade ligeiramente mais restrita do que a presente no
MATLAB. As contribuições de pacotes são centralizadas no
\href{https://octave.sourceforge.io/}{Octave Forge}, um projeto colaborativo
dedicado a expandir as funcionalidades do GNU Octave, por meio de uma seleção
abrangente de pacotes e ferramentas. Essa coleção engloba uma ampla variedade
de domínios, incluindo processamento de sinais, processamento de imagens,
otimização, estatísticas, sistemas de controle, e muito mais.

Octave é uma linguagem interpretada que suporta diferentes paradigmas de
programação, dentre eles a programação procedural, a programação vetorial e a
programação orientada a objetos.  Além disso, a linguagem é dinamicamente
tipada, ou seja, não precisa declarar explicitamente o tipo de uma variável,
sendo este definido automaticamente durante a execução do programa.  A
linguagem suporta tipos de dados usuais como ponto flutuante, inteiros,
\emph{strings}, matrizes e estruturas. O Octave utiliza ainda estruturas de
controle de fluxo padrão, como \emph{loops} (\emph{for}, \emph{while}) e condicionais
(\emph{if-else}).  Podemos ainda definir funções (inclusive funções anônimas) e
utilizar expressões de índice, expressões para realizar operações aritméticas,
comparações e atribuições.  É possível ainda tratar erros e avisos, bem como
depurar um código.

\section{Tipos básicos de dados} 

\epigraph{Às vezes me acontecia acreditar em seis coisas impossíveis antes do
café da manhã.\footnotemark} {Lewis Carroll, Alice Através do Espelho}
\footnotetext{Texto original: \emph{Why, sometimes I’ve believed as many as six
impossible things before breakfast}.}

Nas linguagens de programação, os dados são representados internamente por
diferentes tipos, cada um com suas próprias características e limitações. No
GNU Octave, os tipos básicos de dados podem ser divididos em: numéricos,
\emph{strings}, lógicos, células e estruturas. Embora certas operações
vetoriais possam ser aplicadas também a tipos lógicos e \emph{strings}, o foco
aqui será nos elementos numéricos. A representação binária de números pode
gerar resultados contraintuitivos devido à precisão limitada. Em operações
repetitivas, pequenos erros de precisão podem se acumular e causar grandes
desvios no resultado final. Por exemplo, somar um milésimo mil vezes pode não
resultar exatamente em 1 devido a essas limitações. Além disso, operações com
números muito grandes podem levar a \emph{overflow}\footnote{Um \emph{overflow}
ocorre quando uma operação aritmética gera um resultado fora da faixa
representável pelo tipo numérico utilizado.} (ou saturação, caso existam
mecanismos para preveni-lo) e, ao lidar com números muito pequenos, podemos
incorrer em \emph{underflow}\footnote{Um \emph{underflow} ocorre quando uma
operação aritmética resulta em um número tão pequeno que a precisão da
representação adotada não é suficiente para representá-lo corretamente.}.

O tipo numérico padrão é \emph{double}, dupla precisão no formato de ponto
flutuante. Trata-se de um formato de representação de 64 bits definido no
padrão IEEE 754 \cite{ieee754}. Adota-se este como padrão para manter maior acurácia nos
cálculos numéricos realizados computacionalmente e também por poderem
representar uma larga gama de valores, desde valores muito grandes a valores
muito pequenos, sem que se tenha perda de precisão significativa. Apesar de ser
uma boa forma de representar computacionalmente números e realizar cálculos, a
\Cref{lst-binrep} apresenta um exemplo de como as limitações inerentes à
representação binária por ponto flutuante pode nos levar a resultados
inesperados\footnote{Para contornar as limitações impostas pela aritmética de
ponto flutuante, existem bibliotecas para realizar calculo aritmético de
precisão arbitrária (também utiliza-se o termo \emph{bignums} para referenciar
tal representação), sendo limitado apenas pela memória disponível no sistema.}.
Devemos considerar que estamos suscetíveis a erros nas representações binárias
dos números.  Neste exemplo, a diferença entre o valor previsto e o valor
calculado é maior do que o épsilon de máquina (\lstinline{eps}).

\begin{lstlisting}[language=octave, escapechar=|, label=lst-binrep, caption={A imprecisão inerente à representação binária pode nos levar a resultados inesperados. Observe como a soma de duas frações \sfrac{1}{10} e \sfrac{2}{10} nos leva a um resultado diferente do esperado \sfrac{3}{10}.}]
|\myOctave|1/10 + 2/10 == 3/10
ans = 0
|\myOctave|eps
ans = 2.2204e-16
\end{lstlisting}

Estamos sujeitos aos erros representacionais inerentes à representação binária com
precisão finita e também a erros advindos da imprecisão de cálculos aritméticos computacionais.
Veja na \Cref{lst-binarith} como, mesmo utilizando números bem representados na forma binária,
podemos ter imprecisões provenientes da operação numérica. 
\begin{lstlisting}[language=octave, escapechar=|, label=lst-binarith, caption={Imprecisão oriunda da implementação numérica da divisão.}]
|\myOctave|a = 0.7777777777777;   |\myOctave|b = 7;                 |\myOctave|c = 0.1111111111111;   |\myOctave|a/b == c
ans = 0
\end{lstlisting}

Ao definir uma variável atribuindo a ela um valor numérico, será criada uma
variável matricial com elementos do tipo \lstinline{double}. Mesmo que o intuito
seja representar um escalar, ainda assim o mesmo será representado por uma
matriz, porém de dimensões $1\times 1$. Qualquer variável de valor numérico no
GNU Octave será então representada por uma matriz. Algumas operações numéricas
são integradas através de funções codificadas em \CC{} e compiladas,
fornecendo assim performance e eficiência na realização de cálculos numéricos.
Dentre elas podemos citar as operações aritméticas básicas que são
implementadas através de funções cujo nome remete à operação em questão. Por
exemplo, a função \lstinline{plus} é responsável pela operação da adição
matricial, sendo sua sintaxe dada por \lstinline{plus (X1, X2, ...)}, entretanto é
mais prático e intuitivo utilizar a forma através da sobrecarga de operador\footnote{A
    sobrecarga de operadores é um tipo específico de polimorfismo em que os
    operadores têm implementações diferentes para diferentes argumentos. Esse
    conceito permite aos programadores utilizar uma notação mais próxima do
    domínio em que o código está sendo empregado, tornando-o mais expressivo e
    legível.}: \lstinline{X1 + X2 + ...}.
Da mesma forma, a função \lstinline{minus} implementa a subtração, sendo possível
também executá-la através do operador \lstinline{-}, na forma \lstinline{X1 - X2 - ...}. A lista completa pode ser
acessada na \href{https://docs.octave.org/latest/Arithmetic-Ops.html}{documentação}
do Octave \cite{octave}.

O trecho apresentado na \Cref{lst-plus} mostra como a função \lstinline{plus} é
utilizada para somar vetores, sendo executada de forma significativamente mais
rápida do que a sua contraparte realizada através de um loop para somar
elemento a elemento dos vetores. No teste em questão, observou-se que a
utilização da função integrada foi $470\times$ mais rápida.

\begin{lstlisting}[language=octave, escapechar=|, label=lst-plus, caption={Exemplo de execução da função \lstinline{plus} para realizar a adição de dois vetores. Utilizou-se diretamente a função (linha \ref{l-plus}), através de sobrecarga de operador (linha \ref{l-plus2}) e um loop para realizar a adição elemento a elemento (linha \ref{l-plus-loop}).}]
|\myOctave|X = [1:1E8];
|\myOctave\label{l-plus}|tic; plus (X, X); toc;
Elapsed time is 0.585023 seconds.
|\myOctave\label{l-plus2}|tic; X + X; toc;
Elapsed time is 0.587963 seconds.
|\myOctave\label{l-plus-loop}|tic; for i = 1:length (X), X(i) + X(i); endfor; toc;
Elapsed time is 274.972 seconds.
\end{lstlisting}

Além do ganho em desempenho na execução, observe como a notação através do
operador \lstinline{+} é mais intuitiva e próxima à notação matemática. Tal
abstração elimina os detalhes subjacentes sobre como a operação de adição sobre
vetores é internamente implementada. Este processo é essencial para lidar com
problemas complexos, uma vez que a infinidade de detalhes ofuscaria nossa
real compreensão sobre o problema.
 
\section{Notação como ferramenta de introspecção}\label{sec-notation}

\epigraph{Os limites da minha língua são os limites do meu mundo.\footnotemark}
{Ludwig Wittgenstein, Tractatus Logico-Philosophicus} \footnotetext{Texto
original: \emph{Die Grenzen meiner Sprache bedeuten die Grenzen meiner Welt}.}

\textcite{iverson1980} propõe que a notação utilizada na programação deve
funcionar também como uma ferramenta para a introspecção, de forma a facilitar
a compreensão e a análise dos algoritmos, tornando-se uma forma propícia para
expressar a problemática abordada na resolução de um problema. Ele priorizou a
criação de uma notação que fosse concisa, expressiva e próxima da notação
matemática utilizada em problemas matemáticos e científicos. Criou assim a
linguagem de programação APL (\emph{A Programming Language}) da qual o GNU
Octave pode ser considerado um descendente. O Octave, sendo uma linguagem de
programação voltada para cálculos científicos e numéricos, se alinha a essa
proposta ao utilizar uma notação próxima à notação da matemática utilizada na
álgebra linear. Como vimos anteriormente na \Cref{lst-plus}, o Octave permite
a realização de operações matriciais e vetoriais de maneira concisa e
eficiente. Veremos a seguir mais alguns paralelos entre essa linguagem e a
notação matemática. 

A definição de um vetor em Octave assemelha-se à representação algébrica.
A seguir apresentamos a notação matemática para definir um vetor:
$X = \big[\begin{smallmatrix} 
1 & 2 & 3 & 4 & 5 & 6 & 7 & 8 & 9 
\end{smallmatrix}\big]$.
De forma semelhante, a notação em Octave para definir o mesmo vetor é \lstinline{X = [1 2 3 4 5 6 7 8 9]}.
Para definir a soma dos elementos do vetor, matematicamente expressamos através de
$S = \sum_{i=1}^{|X|} x_i$, onde utilizamos $|X|$ para expressar o número de elementos no vetor $X$
e $x_i$ o $i$-ésimo elemento deste vetor. No Octave a mesma soma pode ser representada por
\lstinline{S = sum (X)}, que implicitamente equivale a realizar um loop somando os elementos de \lstinline{X},
porém a função pré-definida\footnote{
    Existem diversas funções pré-definidas no Octave. Aquelas mais básicas, como é o caso da função 
    função \texttt{sum}, são implementadas em \CC{} e compiladas, sendo assim computacionalmente mais eficientes
    que uma função implementada em script para Octave. Para saber como é a implementação de uma função vejo o \emph{help}
    da mesma. A seguir apresentamos a primeira linho \emph{help} da função \texttt{sum}: 
    \texttt{'sum' is a built-in function from the file libinterp/corefcn/data.cc}.
    Note que há referência ao arquivo \texttt{data.cc}, contendo o código dessa função.
    Mesmo que seja uma função pré-definida, implementada diretamente na linguagem do Octave, os programadores
    que criam tais funções têm o cuidado de criar um código eficiente. 
    Em geral, vale a regra: sempre que possível utilize as funções pré-definidas.
}
\lstinline{sum} é mais eficiente.
Note ainda que a representação anterior é mais intuitiva do que 
a representação através de um loop: \lstinline{S = 0; for i = 1:length (X), S += X(i); endfor}.
No exemplo anterior, utilizamos algumas expressões sintáticas que não
estão disponíveis no MATLAB, dentre elas o operador \lstinline{+=} e
a utilização da instrução \lstinline{endfor} para fechar o escopo do loop \lstinline{for}.
O GNU Octave introduziu ainda outras pequenas diferenças em sua sintaxe,
em relação à linguagem do MATLAB. Uma breve descrição delas pode ser lida na 
\href{https://wiki.octave.org/wiki/index.php?title=Differences_between_Octave_and_Matlab}{GNU Octave Wiki} \cite{octavewiki}.

A resolução de um sistema linear de equações pode ser representada
matematicamente por: $Ax=b$, onde $A$ é a matriz de coeficientes, $x$ é o vetor
de incógnitas e $b$ é o vetor de constantes conhecidas. O Octave utiliza a
notação \lstinline{\}, operador de divisão à esquerda, para expressar a
resolução de sistema lineares de forma concisa e eficiente. Aplicando este
operador podemos resolver o sistema linear encontrando assim os valores das
incógnitas: \lstinline{x = A \ b}. O operador \lstinline{\} equivale à função
\lstinline{mldivide}. A lista completa de funções e operadores sobrecarregados
pode ser vista na
\href{https://docs.octave.org/latest/Operator-Overloading.html}{documentação}
do Octave \cite{octave}.

Centenas de anos de desenvolvimento moldaram a notação matemática utilizada
atualmente \cite{cajori2003}. Essa notação desempenhou um papel preponderante nos avanços
científicos ao longo da história, servindo como uma língua franca para a
comunicação e expressão de conceitos matemáticos complexos. De maneira similar,
as ferramentas computacionais disponíveis podem adotar notações distintas e
também estão em constante evolução. Cada problema e sua solução podem ser
expressos de maneira mais eficaz através de linguagens de programação que
adotam paradigmas adequados para representá-los. Como discutimos anteriormente,
o GNU Octave, por exemplo, foi projetado para ser uma ferramenta conveniente à
resolução de problemas de álgebra linear e lidar com matrizes, tornando a
interação com o mundo da matemática e da computação mais eficaz e intuitiva.

Nas próximas seções iremos apresentar alguns aspectos da notação vetorial 
no Octave. Através de exemplos, ilustraremos como a utilização dessa notação
aproxima-se da representação matemática, evidenciaremos o ganho de desempenho
ao se adotar tal notação para a qual o Octave foi otimizado e ainda como 
a utilização de tal notação torna nosso código menos susceptível a erros.

  \subsection{Algebra Linear}\label{sec-algebralinear} 

A Álgebra Linear desempenha um papel fundamental nas ciências, sendo uma
linguagem universal para descrever e resolver uma ampla gama de problemas nas
áreas de engenharia, física, ciência da computação, estatísticas, dentre
outras \cite{strang2005}.

Octave foi criado para abordar problemas complexos de Álgebra Linear de forma
eficiente e intuitiva. Um dos principais fatores que contribuem para tanto é a
utilização de uma notação que se assemelha à notação matemática utilizada em
Álgebra Linear. Dessarte, o Octave funciona como uma ferramenta para a
introspecção sobre a representação matemática de problemas, sendo possível
representar conceitos diretamente em sua linguagem, sem a necessidade de
conversões complexas entre a notação matemática e sua implementação
computacional.

A fim de ilustrar alguns aspectos, considere dois vetores em $\mathbb{R}^n$:
$\mathbf{a} = [a_1, a_2, \cdots, a_n]^\intercal$ e $\mathbf{b} = [b_1, b_2, \cdots, b_n]^\intercal$.
O produto escalar entre eles é dado por
\begin{equation*}
\mathbf a \cdot \mathbf b = \mathbf a^\intercal \mathbf b =  \sum_{i=1}^n a_i b_i = a_1 b_1 + a_2 b_2 + \cdots + a_n b_n .
\end{equation*}
Na \Cref{lst-prod-escalar} apresentamos o cálculo deste produto escalar.  Na
linha \ref{ln-erro} cometemos o erro de não definir previamente a variável
\lstinline{r}, na qual acumularemos a soma das parcelas $a_i b_i$ para cada $i$
de $1$ a $n$. Já a notação adotada na linha \ref{ln-prodsc} reflete a notação
matemática $\mathbf a^\intercal \mathbf b$, sendo assim também mais compacta e
por isso acaba tornando-se menos suscetível a erros em sua implementação. Note
também (veja as linhas \ref{ln-prdtm} e \ref{ln-prdsctm}) como a execução
usando a notação vetorial é aproximadamente $700\times$ mais rápida.

\begin{lstlisting}[language=octave, escapechar=|, label=lst-prod-escalar, caption={Produto escalar de dois vetores.}]
|\myOctave|a = [1 2 3]'; b = [3 2 1]';
|\myOctave\label{ln-erro}|for i = 1:length (a), r += a(i)*b(i); endfor
error: in computed assignment A OP= X, A must be defined first
|\myOctave|r = 0; for i = 1:length (a), r += a(i)*b(i); endfor
|\myOctave\label{ln-prodsc}|r = a'*b;
|\myOctave\label{ln-defarand}|a = rand (1E6,1); b = rand (1E6,1);
|\myOctave\label{ln-prdtm}|tic; r = 0; for i = 1:length (a), r += a(i)*b(i); endfor; toc;
Elapsed time is 2.76726 seconds.
|\myOctave\label{ln-prdsctm}|tic; r = a'*b; toc;
Elapsed time is 0.00390911 seconds.
\end{lstlisting}

Na \Cref{sec-pca} apresentaremos a análise em componentes principais como
pano de fundo para novamente abordar a notação vetorial, mostrar sua proximidade
com a notação matemática e como tal abordagem propicia uma maior intuição 
sobre o problema.

   \subsubsection{Análise em Componentes Principais}\label{sec-pca}

Para ilustrar como a notação vetorial matemática assemelha-se à notação
utilizada no Octave, vamos apresentar como exemplo a Análise em Componentes
Principais (PCA, \emph{Principal Components Analysis}). A PCA é uma
transformação linear ortogonal que permite projetar dados em um novo sistema de
coordenadas, no qual a maior parte da variância dos dados é representada em
poucas coordenadas. Isso significa que a primeira dimensão conterá a maior
parte da variância e será chamada de componente principal, enquanto a segunda
dimensão conterá a segunda maior parte da variância, e assim por diante.

Para realizar essa transformada, devemos encontrar a matriz dos autovetores da
matriz de correlação dos dados. Seja $\mathbf{X}$ a matriz $m \times n$, com as
amostras dos dados em $\mathbb{R}^n$ dispostos ao longo das linhas de
$\mathbf{X}$, então a matriz de autocorrelação de $\mathbf{X}$ é dada por
\begin{equation*}\label{eq-covmtx}
 \mathbf{S}_{\mathbf{X}} = \frac{\mathbf{X} \mathbf{X}^\intercal}{n-1}.
\end{equation*}
A matriz ortogonal $\mathbf{P}$ que faz a transformação da PCA é a matriz em que os
autovetores de $\mathbf{S}_{\mathbf{X}}$ dispostos como as linhas de $\mathbf{P}$.
Ao aplicar a transformação, ou seja, projetar os dados nas linhas de $\mathbf{P}$,
obteremos a matriz $\mathbf{Y}$,
\begin{equation*}\label{eq-pca}
 \mathbf{Y} = \mathbf{P}^\intercal \mathbf{X}.
\end{equation*}
A tradução para o Octave é simples, como ilustrado na \Cref{lst-pca}. Isso evidencia
a similaridade da notação matemática com a notação vetorial utilizada no Octave,
facilitando assim a introspecção sobre a transformada através da análise de sua
implementação nessa linguagem.

Um passo importante de pré-processamento antes da aplicação da PCA é a subtrair
a média ao longo de cada dimensão dos dados. Tal passo é importante para
eliminar qualquer viés ou deslocamento que possa existir nos dados. Isso
garante que as direções da PCA representem adequadamente a estrutura de
variação dos dados, sem serem afetadas por deslocamentos nos dados. Essa etapa
é implementada através do trecho \lstinline{X -= mean (X, 2)} no código da
\Cref{lst-pca}. Essa forma de implementação torna conciso, intuitivo e eficiente esse
passo de normalização. Para uma maior concisão e eficiência, tal notação ainda faz
uso do \emph{broadcast} de operador, conceito que será visto na \Cref{sec-broadcast}. 
Calcular a média para cada dimensão e, através de um
loop, normalizar a média seria mais complexo, o que poderia ser implementado
por: \lstinline{mm = mean (X, 2); for i = 1:length (mm), data(i,:) -= mm(i); endfor}. 

\begin{lstlisting}[language=octave, escapechar=|, label=lst-pca, caption={Análise em Componentes Principais. Definição simplificada da função e um exemplo de utilização.}]
|\myOctave|function Y = pca (X),
> n = columns (X);
> X -= mean (X, 2);
> S = X * X' / (n-1);
> [P, _] = eig(S);
> Y = P'*X;
> endfunction
|\myOctave|theta = pi/4;
|\myOctave|R = [cos(theta) -sin(theta); sin(theta)  cos(theta)];
|\myOctave|X = R*(randn (2, 100) .* [1; 0.1]);
|\myOctave|Y = pca (X);
|\myOctave|figure; subplot (1,2,1); plot (X(1,:), X(2,:), '.'); subplot (1,2,2); plot (Y(1,:), Y(2,:), '.');
\end{lstlisting}

  \subsection{Indexação}\label{sec-idx}

Expressões de índice permitem referenciar elementos em uma matriz a partir de
índices.  Os índices podem ser escalares, vetores ou intervalos. Podemos
utilizar ainda o operador especial `\lstinline{:}' para expressar uma
linha/coluna completa ou para descrever um intervalo. Por exemplos,
\lstinline{2:8} descreve o intervalo de dois a oito ($[2,8]$).  Podemos ainda
utilizar a palavra-chave `\lstinline{end}' para designar o último elemento,
eliminando assim a necessidade de chamar as funções \lstinline{size} ou
\lstinline{length}.

Uma expressão de índice consiste em um par de parênteses \lstinline{( ... )}
cercando $n$ expressões separadas por vírgulas, onde cada uma dessas expressam
um índice associado a cada uma das $n$ dimensões do objeto. É possível ainda
utilizar um índice linear para indexar os elementos de uma matriz. Nesse caso,
devemos lembrar que os elementos são organizados primeiramente ao longo do
primeiro índice (linha), depois ao longo do segundo (colunas) e assim por
diante. 

Para exemplificarmos a utilização de expressões de índice, considere a matriz
mágica\footnote{Matriz mágica, ou quadrado mágico, é a denominação dada à
    matriz quadrada em que a soma das linhas, colunas ou diagonais principais
    são iguais. Note que a matriz do exemplo dado tem como resultado de
qualquer uma das somas o valor 34.} bidimensional \lstinline{M} de tamanho $4
\times 4$ definida na \Cref{lst-magic-square}.

\begin{lstlisting}[language=octave, escapechar=|, label=lst-magic-square, caption={Matriz mágica utilizada nos exemplos.}]
|\myOctave|M = magic (4);
|\myOctave|disp (M)
   16    2    3   13
    5   11   10    8
    9    7    6   12
    4   14   15    1
\end{lstlisting}

Podemos indexar os elementos de uma matriz através de um índice linear, como
exemplificado nas linhas \ref{ex-ind1}, \ref{ex-ind2} e \ref{ex-ind3} ou
através de dois índices (linha e coluna), como ilustrado nas linhas
\ref{ex-ind4}, \ref{ex-ind5} e \ref{ex-ind6}.

\begin{multicols}{2}[\captionof{lstlisting}{Exemplos de utilização de expressões de índice.}]
\begin{lstlisting}[language=octave, escapechar=|, label=lst-exp-index]
|\myOctave\label{ex-ind1}|disp (M(3:7))
    9    4    2   11    7
|\myOctave\label{ex-ind2}|disp (M(12:end))
   15   13    8   12    1
|\myOctave\label{ex-ind3}|disp (M([1,3,5]))
   16    9    2
|\columnbreak|
|\myOctave\label{ex-ind4}|disp (M(1:2,2:3))
    2    3
   11   10
|\myOctave\label{ex-ind5}|disp (M(end,end-1:end)
   15    1
|\myOctave\label{ex-ind6}|disp (M(3,:))
    9    7    6   12
\end{lstlisting}\end{multicols}

Utilizar expressões de índice facilita a manipulação de dados, permite utilizar
operações vetoriais sobre elementos de matrizes, torna ainda o código mais
legível, eficiente e conciso.  Veja mais sobre
\href{https://docs.octave.org/latest/Index-Expressions.html}{expressões de
índice} na documentação do Octave.

  \subsubsection{Indexação Lógica}\label{sec-idx-logic}

Uma outra forma de indexação muito útil é a indexação lógica.
Neste tipo de indexação, utilizamos um vetor lógico para extrair
os valores não nulos correspondentes a este vetor lógico. 

Como exemplos, vamos novamente utilizar a matriz mágica definida na 
\Cref{lst-magic-square}.  
Suponhas que desejamos extrair os elementos de \lstinline{M} menores do que $8$.
Na linha \ref{l-idxM} podemos verificar que a expressão \lstinline{M < 8} retorna
$1$ (um) para os elementos que satisfazem tal condição e $0$ (zero) para os demais.
Podemos utilizar este vetor de índices lógicos para extrair os elementos desejados,
assim como está na linha \ref{l-idxM2}.

\begin{multicols}{2}[\captionof{lstlisting}{Exemplo de utilização da indexação lógica.}]
\begin{lstlisting}[language=octave, escapechar=|, label=lst-logic-index]
|\myOctave\label{l-idxM}|disp (M < 8)
  0  1  1  0
  1  0  0  0
  0  1  1  0
  1  0  0  1
|\columnbreak|
|\myOctave\label{l-idxM2}|disp (M(M < 8)')
   5   4   2   7   3   6   1
\end{lstlisting}\end{multicols}

Diversas funções do Octave, cujos nomes começam com \lstinline{is}, retornam um vetor lógico.
Por exemplo, suponha que desejamos substituir os elementos \lstinline{NaN} de uma matriz por $0$ (zero).
Utilizando novamente a matriz mágica \lstinline{M} como exemplo, na \Cref{lst-logic-index-nan}, iremos introduzir
uma \lstinline{NaN} em seu primeiro elemento e depois substituí-lo por $0$ (zero). Utilizamos para tanto
a função \lstinline{isnan}, que retorna uma matriz de índice lógico com entradas nulas para os elementos
que não são \lstinline{NaN} e $1$ (um) para os elementos \lstinline{NaN}.

\begin{multicols}{2}[\captionof{lstlisting}{Utilização da função \lstinline{isnan} para remover os elementos \lstinline{NaN}.}]
\begin{lstlisting}[language=octave, escapechar=|, label=lst-logic-index-nan]
|\myOctave|M(1,1) = 0/0;
|\myOctave|disp (isnan (M))
  1  0  0  0
  0  0  0  0
  0  0  0  0
  0  0  0  0
|\myOctave|M(isnan (M))=0;
|\myOctave|disp (M)
    0    2    3   13
    5   11   10    8
    9    7    6   12
    4   14   15    1
\end{lstlisting}\end{multicols}

Utilizar indexação lógica, além de prático, é mais eficiente. Para atestar, na
\Cref{lst-logic-index2}, vamos criar um vetor aleatório com inteiros entre $1$
e $100$ para calcular a média daqueles com valor maior do que $50$. Sem
utilizar indexação lógica, devemos realizar um \emph{loops} pelos elementos do vetor,
verificar se cada elemento é maior do que $50$, somar, contabilizar quantos
elementos encontramos e então dividir a soma pelo número de elementos (veja
linha \ref{l-mean50loop}). Na linha \ref{l-mean50} obtemos o mesmo resultado,
porém utilizando indexação lógica. Note a simplicidade e eficiência. Para um
vetor com um milhão de números, o código utilizando indexação lógica foi
praticamente $200\times$ mais rápido.

\begin{lstlisting}[language=octave, escapechar=|, label=lst-logic-index2, caption={Teste de tempo de execução para evidenciar a eficiência em se utilizar a indexação lógica.}]
|\myOctave|r = randi(100, 1, 1E6);
|\myOctave\label{l-mean50loop}|tic; s = 0; c = 0; for i=1:length (r), if(r(i)>50) s+=r(i); c++; endif; endfor; m = s/c; toc;
Elapsed time is 2.29225 seconds.
|\myOctave\label{l-mean50}|tic; mean ( r(r > 50) ); toc;
Elapsed time is 0.0117271 seconds.
\end{lstlisting}

Duas funções uteis para se utilizar com indexação lógica são as funções \lstinline{any} e
\lstinline{all}. Elas são utilizadas para determinada se algum ou todos, respectivamente,
elementos de uma matriz satisfazem determinada condição. Por exemplo, na \Cref{lst-logic-index3},
utilizamos tais funções para gerar avisos caso algum elemento de um vetor seja menor
do que $10$ ou caso todos elementos sejam maiores que $40$.

\begin{lstlisting}[language=octave, escapechar=|, label=lst-logic-index3, caption={Exemplos de utilização das funções \lstinline{any} e \lstinline{all}.}]
|\myOctave|if any (r < 10), warning ("Existe valor menor que 10!"); endif
|\myOctave|if all (r > 40), warning ("Todos valores são maiores que 40!"); endif
warning: Todos valores são maiores que 40!
\end{lstlisting}

   \subsubsection{Varredura dos elementos de uma matriz}\label{sec-mscan}
Com a finalidade de ilustrar como a notação vetorial e a utilização de 
funções pré-definidas facilitam o entendimento, a resolução do problema
e ainda resultam em um código mais eficiente, mostraremos como estes 
conceitos podem ser aplicados ao problema de varredura dos elementos
de uma matriz.

\begin{figure}[htbp]
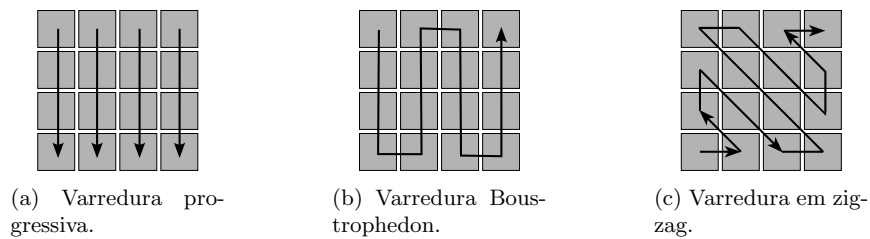

\centering
\begin{minipage}[t]{0.25\textwidth}
\centering
\includegraphics[width=0.75\textwidth]{scan_lin.pdf}
\subcaption{Varredura progressiva.}\label{img-linearscan}
\end{minipage}
\hfill
\begin{minipage}[t]{0.25\textwidth}
\centering
\includegraphics[width=0.75\textwidth]{scan_prog.pdf}
\subcaption{Varredura Boustrophedon.}\label{img-boustrophedon}
\end{minipage}
\hfill
\begin{minipage}[t]{0.25\textwidth}
\centering
\includegraphics[width=0.75\textwidth]{zigzag4x4.pdf}
\subcaption{Varredura em zig-zag.}\label{img-zigzag}
\end{minipage}
\caption{Três tipos de varredura de matriz comumente utilizados (direções alteradas para simplificar).}
\label{fig-imgscan}
\source{Autoria própria.}
\end{figure}

A \Cref{fig-imgscan} apresenta três tipos de varredura de matriz que são muito
utilizados, principalmente no contexto de processamento de imagens digitais. A
varredura linear (\cref{img-linearscan}) é a mais simples onde as
linhas/colunas são lidas sequencialmente. A varredura boustrophedon
(\cref{img-boustrophedon}) é semelhante à linear, porém para evitar
descontinuidade ao passar de uma linha/coluna para a próxima, inverte-se o
sentido de leitura. Por fim, a varredura em zigzag (\cref{img-zigzag}) busca
ordenar os elementos da matriz seguindo um padrão zigzag. Esse último tipo de
varredura é utilizado, por exemplo, no padrão JPEG para ordenar os coeficientes
da transformada em cossenos. Adotamos a leitura de colunas como a forma padrão,
seguindo assim o padrão adotado pelo Octave. Para a leitura em zigzag, adotamos
o ponto inicial como sendo o elemento inferior mais à esquerda e iniciamos o
zigzag partindo para o elemento à direita. Essa escolha foi feita apenas
para simplificar o exemplo aqui ilustrado.

Para exemplificar o processo de varredura, vamos utilizar como exemplo a
matriz mágica $4 \times 4$ dada na \Cref{lst-magic-square}. 
Em uma linguagem de programação estruturada, para realizar a leitura dos
valores de uma matriz seguindo a ordem linear (\cref{img-linearscan}), faríamos
\emph{loops} aninhados para percorrer os elementos da matriz ao longo de suas linhas e
colunas. A \Cref{lst-scan-loop-linear} representa como pode ser implementada
tal estratégia utilizado o Octave.

\begin{lstlisting}[language=octave, escapechar=|, label=lst-scan-loop-linear, caption={Loop para a varredura linear dos elementos de uma matriz.}]
|\myOctave|for j = 1:size (M,2), for i = 1:size (M,1), m((j-1)*size (M,2)+i) = M(i,j); endfor; endfor; disp (m)
   16    5    9    4    2   11    7   14    3   10    6   15   13    8   12    1
\end{lstlisting}

Entretanto, podemos utilizar expressões de índice (veja \Cref{sec-idx}) para
simplificar a varredura linear dos elementos da matriz. 
Veja a implementação apresentada na \Cref{lst-scan-linear},
onde a simples expressão \lstinline{M(:)} resulta em um vetor coluna. Tal
expressão é um `idioma de código'\footnote{Um `idioma de código' ou `idioma de
programação' é um fragmento de código que desempenham um determinado papel
semântico e que se repete com frequência em diversos códigos.} encontrado com
muita frequência, é mais rápido e geralmente mais claro do que realizar um
loop ou ainda chamar a função \lstinline{reshape}.

\begin{lstlisting}[language=octave, escapechar=|, label=lst-scan-linear, caption={Varredura linear dos elementos da matriz utilizando uma simples expressão de índice.}]
|\myOctave|disp (M(:)')
   16    5    9    4    2   11    7   14    3   10    6   15   13    8   12    1
\end{lstlisting}

Na \Cref{lst-scan-loop-prog} apresentamos um exemplo estruturado para a implementação
da varredura boustrophedon dos elementos da matriz. Essencialmente é semelhante à varredura
linear, porém devemos inverter o sentido de leitura nas colunas pares.

\begin{lstlisting}[language=octave, escapechar=|, label=lst-scan-loop-prog, caption={Varredura boustrophedon dos elementos de uma matriz utilizando \emph{loops} aninhados.}]
|\myOctave|for j = 1:size (M,2), for i = 1:size (M,1),
> if (mod (j,2) == 1), v = M(i,j);
> else, v = M(size (M,1)-i+1,j);
> endif;
> m((j-1)*size (M,2)+i) = v;
> endfor; endfor;
|\myOctave|disp (m);
   16    5    9    4   14    7   11    2    3   10    6   15    1   12    8   13
\end{lstlisting}

Este processo de varredura pode ser expresso de forma simplificada utilizado
expressão de índice para selecionar as colunas pares e a função
\lstinline{flipud} para inverter a ordem dos elementos de tais colunas.
Observando o \emph{help} e o código dessa função, constatamos que ela simplesmente
gera um vetor de índices reverso (\lstinline{size (x,dim):-1:1}) 
para indexar os elementos. Dessa forma, as linhas \ref{line:flipup} e \ref{line:invindex} da \Cref{lst-scan-prog}
são equivalentes, entretanto a primeira utiliza um nível extra de abstração,
tornando o código mais intuitivo e sem criar maiores complexidades para lidar
com os índices que referenciam os elementos das colunas pares.

\begin{lstlisting}[language=octave, escapechar=|, label=lst-scan-prog, caption={Varredura boustrophedon dos elementos da matriz utilizando expressão de índice (unicamente ou também através da função \texttt{flipud}).}]
|\myOctave\label{line:flipup}|M(:,2:2:end) = flipud (M(:,2:2:end)); disp (M(:)')
   16    5    9    4   14    7   11    2    3   10    6   15    1   12    8   13
|\myOctave\label{line:invindex}|M(:,2:2:end) = M([size (M,2):-1:1],2:2:end); disp (M(:)')
   16    5    9    4   14    7   11    2    3   10    6   15    1   12    8   13
\end{lstlisting}

O último exemplo proposto é a varredura em zigzag.  Essa forma de varredura é
mais complexa de ser implementada. Vamos começar a partir do elemento mais à
esquerda na parte inferior da matriz e prosseguiremos como ilustrado na
\Cref{img-zigzag}. A primeira resolução, apresentada na
\Cref{lst-scan-zigzag}, utilizamos uma variável \lstinline{c} que define o sentido
da varredura, quando esta é negativa, a varredura prossegue subindo uma
diagonal ou, quando é positiva, descendo uma diagonal. Ao chegar nos limites da
matriz é necessário ajuste dos contadores de posição para seguir percorrendo
seus elementos.

\begin{lstlisting}[language=octave, escapechar=|, label=lst-scan-zigzag, caption={Varredura em zigzag da matriz utilizando loop e estrutura de seleção.}]
|\myOctave|function r = inrange(i,m), 
> if m(1) <= i && i <= m(2), r = true; 
> else, r = false; 
> endif; 
> endfunction
octave:16> function m = zigzag_loop(M),
> k = 1; c = 1; i = size (M,1); j = 1;
> while (inrange (i, [1 size(M,1)]) && (inrange (j, [1 size(M,2)]))),
>   m(k) = M(i,j); i = i+c; j = j+c; k++;
>   if i < 1 && j < 1, i++; j += 2; c *= (-1); 
>   elseif i > size (M,1) && j > size (M,2), i -= 2; j--; c *= (-1); 
>   elseif i < 1, i++; j += 2; c *= (-1); 
>   elseif i > size (M,1), i--; c *= (-1); 
>   elseif j < 1, j++; c *= (-1); 
>   elseif j > size (M,2), j--; i -= 2; c *= (-1); 
>   endif;
> endwhile
> endfunction
\end{lstlisting}

A função para varredura em zigzag da matriz pode ser simplificada utilizando
para tanto a função \lstinline{spdiags} e a função \lstinline{flipud} (veja a
\Cref{lst-zigzag}). A primeira é utilizada para extrair as diagonais da matriz,
já a segunda é utilizada para alternadamente inverter o sentido de leitura, ora
subindo, ora descendo ao longo das diagonais. Tal implementação é mais concisa
e compreensível devido ao nível de abstração mais elevado empregado.

\begin{lstlisting}[language=octave, escapechar=|, label=lst-zigzag, caption={Varredura em zigzag da matriz utilizando as funções \texttt{spdiags} e \texttt{flipud}.}]
|\myOctave|function m = zigzag(M),
> ind = reshape (1:numel(M), size (M));
> ind = spdiags (ind);
> ind(:,2:2:end) = flipud (ind(:,2:2:end));
> ind(ind==0) = [];
> m = M(ind)';
> endfunction
|\myOctave|disp (zigzag (M))
   32   24   35   19   32   37    4   30   31   48    9   32   39   40   46   35
\end{lstlisting}

Por fim, iremos comparar o tempo de execução da varredura zigzag para as duas
implementações propostas. O resultado é apresentada na \Cref{lst-zigzag-time}.
Note como a segunda implementação é aproximadamente 6 vezes mais rápida.

\begin{lstlisting}[language=octave, escapechar=|, label=lst-zigzag-time, caption={Comparação do tempo de execussão das duas abordagens para varrer a matriz em zig-zag.}]
|\myOctave|tic; for i=1:1E3, m=zigzag_loop (M); end; toc;
Elapsed time is 2.39824 seconds.
|\myOctave|tic; for i=1:1E3, m=zigzag (M); end; toc;
Elapsed time is 0.400645 seconds.
\end{lstlisting}  \subsection{Broadcast}\label{sec-broadcast}

O \emph{broadcasting} é um conceito que se refere à capacidade de realizar
operações entre matrizes de distintos tamanhos de maneira coerente e intuitiva.
Em termos simples, dizemos que os arranjos menores são transmitidos através do
arranjo maior, aplicando um dado operador ao longo do percurso, até terem
dimensões compatíveis. Por exemplo, se temos uma matriz $X_{3 \times 3}$ e um
vetor $Y_{1 \times 3}$, ao fazer \lstinline{X + Y}, teremos que o vetor
\lstinline{Y} será `copiado'\footnote{A implementação interna do
\emph{broadcasting} faz a reutilização dos elementos da matriz de menor
dimensionalidade, sem a necessidade de criar cópia de seus elementos.} ao longo
da outra dimensão para que seja computado \lstinline{X + Y}. Ou seja, quando
uma das dimensões é igual a 1, a matriz com essa dimensão única é replicada ao
longo da outra dimensão até que corresponda à dimensão da matriz com a qual se
está realizado a operação.

\begin{lstlisting}[language=octave, escapechar=|, label=lst-broadcast, caption={Exemplo de aplicação do \emph{broadcast} sobre o operador de adição.}]
|\myOctave|X = [1 2 3; 4 5 6; 7 8 9];
|\myOctave|Y = [10 20 30];
|\myOctave|disp (x + y)
   11   22   33
   14   25   36
   17   28   39
\end{lstlisting}

Exemplos mais simples de \emph{broadcasting} são as operações realizadas com
escalares. Ao realizar a soma \lstinline{X + 10}, a comparação \lstinline{X > 5} ou
\lstinline{min (X, 5)}, estamos utilizando \emph{broadcasting} com escalares.

Um exemplos de aplicação com maior dimensionalidade é a conversão de uma imagem
RGB em tons de cinza. Uma imagem RGB é uma matriz tridimensional em que suas
dimensões são linhas, colunas e componentes de cor. Tratando-se de uma imagem RGB,
a terceira dimensão tem comprimento igual a 3. Uma imagem em tons de cinza pode
ser representada por uma matriz bidimensional, onde usualmente o valor de cada
pixel representa um tom de cinza ou, no caso aqui em questão, o valor de
luminância associado ao pixel. Iremos realizar a conversão das cores em tons de
cinza utilizando a contribuição de luminância relativa proporcionada por cada
uma das componentes RGB, as cores primárias vermelho, verde e azul. A função
\lstinline{rgb2gray} utiliza a conversão conforme dado na Rec.ITU-R BT.601-7
\cite{rbt601-7}. Segundo tal recomendação, a luminância $Y'$ é dada por
\begin{equation}\label{eq-rgb2gray} 
    Y' =  0.299 R' + 0.587 G' + 0.114 B',
\end{equation}
onde $Y'$ representa a luminância com correção gamma\footnote{A correção gamma 
é usualmente utilizada para considerar a resposta não linear da visão humana.}
e as variáveis $R'$, $G'$ e $B'$ representam os
valores das componentes de cor também com correção gamma. Para simplificar,
vamos ignorar a correção gamma e ver como podemos realizar tal cálculo
utilizando \emph{broadcast}. A implementação deste exemplo aparece na
\Cref{lst-img-gray}.

\begin{lstlisting}[language=octave, escapechar=|, label=lst-img-gray, caption={Utilização do \emph{broadcasting} para realizar a conversão de uma imagem colorida em tons de cinza.}]
|\myOctave|url = "https://sipi.usc.edu/database/download.php?vol=misc&img=4.1.05";
|\myOctave|urlwrite (url, "house.tiff");
|\myOctave|img = imread ("house.tiff");
|\myOctave|img_gray = sum(img .* permute ([0.299, 0.587, 0.114], [1, 3, 2]), 3);
|\myOctave|figure; subplot (1,2,1); imshow (img); title ('original'); subplot (1,2,2); imshow (uint8 (img_gray)); title ('grayscale');
\end{lstlisting}

Note no exemplo dado que utilizamos a função \lstinline{permute} para permutar
as dimensões da matriz com os coeficientes de peso dados na \Cref{eq-rgb2gray},
ou seja, aplicamos uma generalização da transposição\footnote{A transposição de
matriz é uma simples permutação da primeira e segunda dimensão, ou seja,
\lstinline{permute (A, [2 1])}. Equivale então a aplicar o operador de transposição à matriz
\lstinline{A}, sendo portanto equivalente a \lstinline{A'}.} para matrizes com
mais de duas dimensões.

    \subsubsection{Distância Euclidiana}

Nesta secção apresentamos três implementações de funções para calcular a
distância euclidiana entre todos os pontos de um conjunto. Os pontos são
representados por linhas da matriz \lstinline{P}.

A \Cref{lst-distance1} apresenta a implementação utilizando o paradigma da
programação estruturada. Nesta implementação, realizamos dois \emph{loops} para 
selecionar os pares de pontos (linhas da matriz) e então um outro \emph{loops} 
para calcular a distância entre estes pontos. Note que o segundo loop, para
seleção dos pontos, inicia-se em \lstinline{j = i}, aproveitando assim o fato
de que distâncias possuem a propriedade da simetria, algo que não poderíamos
aplicar se estivéssemos calculando uma divergência, por exemplo a divergência
de Kullback-Leibler.

\begin{lstlisting}[language=octave, escapechar=|, label=lst-distance1, caption={Implementação estruturada utilizando três \emph{loops} para calcular a distância entre todos os pares de pontos.}]
|\myOctave|function d = distance1 (P),
>  [N, M] = size (P);
>  for i = 1:N, for j = i:N,
>    dist = 0;
>    for k = 1:M,
>      dist += (P(i,k) - P(j,k)).^2;
>    endfor;
>    d(i,j) = d(j,i) = sqrt (dist);
>  endfor; endfor
> endfunction
\end{lstlisting}

Na próxima implementação, apresentada na \Cref{lst-distance2}, utilizamos o
\emph{broadcasting} e indexação para calcular a distância de um ponto a todos
demais. Da mesma forma que fizemos anteriormente, ainda utilizaremos a simetria
para simplificar e realizar apenas metade dos cálculos.

\begin{lstlisting}[language=octave, escapechar=|, label=lst-distance2, caption={Implementação utilizando \emph{broadcasting} e indexação para calcular todas as distâncias de um dado ponto aos demais, sendo assim necessários apenas um loop para variar o ponto de referência.}]
|\myOctave|function d = distance2 (P),
>   N = size (P,1);
>   for i = 1:N,
>     d(i:N,i) = sqrt (sum ((P(i:N,:) - P(i,:)).^2,2));
>     d(i,i:N) = d(i:N,i)';
>   endfor
> endfunction
\end{lstlisting}

Por fim, apresentamos uma implementação vetorial que utiliza também o \emph{broadcasting} para calcular todas as 
distância, entre todos os pontos, inclusive os mesmo pares de pontos em ordem reversa, sem aproveitar a simetria do problema.

\begin{lstlisting}[language=octave, escapechar=|, label=lst-distance3, caption={Implementação utilizando \emph{broadcasting} para calcular todas as distâncias de uma vez.}]
|\myOctave|function d = distance3 (P),
>   d = permute (sqrt (sum ((P - permute (P, [3, 2, 1])).^2, 2)), [1,3,2]);
> endfunction
\end{lstlisting}

Na \Cref{lst-distance-time} comparamos o tempo de execução de cada uma das
abordagens para calcular as um milhão de distâncias entre cada par de um
conjunto com mil pontos. Note que a última implementação foi bem superior às
demais, mesmo não fazendo uso da simetria do problema. A primeira
implementação se mostrou a menos eficiente, levando $200 \times$ o tempo de
execução da melhor implementação. Já a segunda apresenta um fator de $30
\times$, ainda significativamente maior que a melhor implementação.

\begin{lstlisting}[language=octave, escapechar=|, label=lst-distance-time, caption={Comparação do tempo de execução de cada uma das funções para calcular as distâncias em um conjunto de mil pontos.}]
|\myOctave|tic; d = distance1 (P); toc;
Elapsed time is 11.0255 seconds.
|\myOctave|tic; d = distance2 (P); toc;
Elapsed time is 1.65677 seconds.
|\myOctave|tic; d = distance3 (P); toc;
Elapsed time is 0.0536449 seconds.
\end{lstlisting}

O \emph{broadcast} permite tornar o código mais curto, inteligível e de fácil
manutenção, além de proporcionar maior eficiência em termos de tempo de
execução e memória. Para demais funções que não são escritas com semântica
propícia ao \emph{broadcasting}, podemos utilizar a função \lstinline{bsxfun}
para coagi-las a aplicar o \emph{broadcasting}. Leia mais sobre o assunto na
\href{https://docs.octave.org/latest/Broadcasting.html}{documentação} do
Octave.

  \subsection{Alças de Funções}\label{sec-functionhandle}

Um \emph{function handle}, ou alça de função, é um objeto que representa uma
função. Isto torna possível atribuir funções a variáveis e passá-las como
argumentos a outras funções. Podemos assim criar funções
anônimas\footnote{Funções anônimas são funções que não estão armazenadas em um
    arquivo de programa, mas estão armazenadas em uma variável do tipo
    \texttt{function\_handle}. Um exemplo simples de definição de uma função
    anônima é dado a seguir: \lstinline{quad = @(x) x.^2;}, onde a variável
\lstinline{quad} armazena um \emph{handle} para uma função que implementa a
função do número quadrado.} para tarefas específicas, passá-las como argumento
a outras funções ou ainda criar funções de ordem superior, que são funções que
operam em outras funções. Dessa forma, podemos trabalhar as funções como
objetos de primeira classe\footnote{No projeto de linguagens de programação, um
objeto de primeira classe é uma entidade que pode aparecer em expressões, ser
atribuída a variáveis e utilizada como argumento.}, permitindo maior
flexibilidade e abstração, podemos criar códigos mais genéricos e
reutilizáveis. 

Como exemplo, na \Cref{lst-rplnaneg}, vamos criar uma função condicional para
substituir valores \lstinline{NaN} ou negativos por zero.

\begin{lstlisting}[language=octave, escapechar=$, label=lst-rplnaneg, caption={Exemplo de função anônima para substituir valores negativos e \lstinline{NaN} por zero.}]
$\myOctave$replace_negative = @(x) (x < 0) .* 0 + (x >= 0) .* x;
$\myOctave$disp (replace_negative ([-1 1 -2 2 -3 3]));
   0   1   0   2   0   3
$\myOctave$replace_neg_nan = @(x) ifelse(isnan(x) | x < 0, 0, x);
$\myOctave$disp (replace_neg_nan ([0 1 2 -1 NaN 3 -2 4]));
   0   1   2   0   0   3   0   4
\end{lstlisting}

\subsubsection{k vizinhos mais próximo}\label{sec-knn}
No exemplo apresentado na \Cref{lst-nn} criamos uma função que utiliza uma
outra função, passada como argumento, para calcular as distâncias entre os
pontos (para cada elemento de \lstinline{Y}, calcula-se as distâncias a todos
elementos de \lstinline{X}). A função retorna o índice do elemento em
\lstinline{X} mais próximo ao correspondente elemento em \lstinline{Y}, e
também a distância entre eles. A função \lstinline{nearestneighbor} recebe como
um de seus parâmetros qual é a função de distância, que deverá ser implementada
como uma \emph{handle} para uma função anônima. No exemplo apresentamos duas
distâncias usuais: distância Euclidiana e distância Manhattan, dadas nas
\Cref{ed-dist-eucl,eq-dist-manh} respectivamente.

\begin{minipage}[m]{0.45\textwidth}
  \begin{equation*}\label{ed-dist-eucl}
      d_e (x,y) = \sqrt{ \sum_{i=1}^{n} (x_i - y_i)^2 },
  \end{equation*}
\end{minipage}
\hfill
\begin{minipage}[m]{0.45\textwidth}
  \begin{equation*}\label{eq-dist-manh}
      d_m (x,y) = \sum_{i=1}^{n} \vert x_i - y_i \vert .
  \end{equation*}
\end{minipage}

\begin{lstlisting}[language=octave, escapechar=|, label=lst-nn, caption={Exemplo de função para retornar o vizinho mais próximo podendo utilizar diferentes distâncias como métrica. Neste exemplos definimos duas funções anônimas para implementar o cálculo da distâncias euclidiana e da distância manhattan.}]
|\myOctave|manhattan = @(X,Y) permute (sum (abs (X - permute (Y, [3 2 1])), 2), [1 3 2]);
|\myOctave|euclidean = @(X,Y) permute (sqrt (sum ((X - permute (Y, [3 2 1])).^2, 2)), [1 3 2]);
|\myOctave|function [idx, d] = nearestneighbor (X, Y, distance)
> > >    D = distance (X, Y);
>    [idx, d] = min (D, [], 2);
> endfunction
|\myOctave|X = [0 0; 1 1; 2 2]; Y = [0 1; 2 1];
|\myOctave|[idx, d] = nearestneighbor (X, Y, manhattan);
|\myOctave|[idx, d] = nearestneighbor (X, Y, euclidean);
\end{lstlisting}

   \subsubsection{Transformada de Cossenos em compressão de imagens}

Algumas funções do Octave necessitam receber como argumento uma função
fornecida pelo usuário. Este é o caso da função \lstinline{blockproc} (do
pacote \emph{image}), apresentada no exemplos da \Cref{lst-dct}. Esta função
aplica a função fornecida a blocos da matriz dada como argumento. No exemplo em
questão, iremos aplicar a DCT\footnote{DCT é o acrônimo para \emph{Discrete
    Cosine Transform} (Transformada Discreta de Cossenos), uma transformada
    ortonormal comumente utilizada em compressão de sinais multimídia, graças à
sua característica de compactação de energia e por lidar apenas com números
reais, quando aplicada a sinais reais.} a blocos de tamanho $8\times 8$ de uma
matriz (este é um dos passos utilizados na compressão JPEG).  Para tanto,
iremos definir a DCT-2D, para matrizes $8\times 8$, através de uma função
anônima que será passada à função \lstinline{blockproc}. Na linha
\ref{l-dctmtx}, definimos a matriz de DCT utilizando a função
\lstinline{dctmtx} (pertencente ao pacote \emph{signal}). Em seguida, na linha
\ref{l-dct2d}, definimos a operação DCT 2D sobre uma matriz (matematicamente é
expressa como $\mathbf{Y} = \mathbf{C} \mathbf{X} \mathbf{C}^\intercal$, onde
$\mathbf{X}$ é a matriz de dados, $\mathbf{C}$ a matriz de DCT e $\mathbf{Y}$ o
resultado de se aplicar a DCT à matriz $\mathbf{X}$). Utilizaremos agora, na
linha \ref{l-blockproc}, a definição da DCT 2D para aplicá-la a blocos de uma
imagem, utilizando para tanto a função \lstinline{blockproc}. Por fim, a
visualização é feita na linha \ref{l-vis-gray}.

\begin{lstlisting}[language=octave, escapechar=|, label=lst-dct, caption={Exemplo de aplicação de função anônima para definir a DCT-2D e utilizá-la em uma das etapas da compressão de imagens.}]
|\myOctave|pkg load signal; pkg load image;
|\myOctave\label{l-blksz}|blk_sz = 8;
|\myOctave\label{l-dctmtx}|T = dctmtx (blk_sz); 
|\myOctave\label{l-dct2d}|dct2d = @(x) T * x * T'; 
|\myOctave\label{l-blockproc}|img_dct = blockproc (img_gray, [blk_sz blk_sz], dct2d);
|\myOctave\label{l-vis-gray}|figure; colormap ('gray'); imagesc (abs (img_dct));
\end{lstlisting}

Através dos exemplos apresentados, observe como a notação adotada contribui
para a eficiência e concisão do código. A leitura do código torna-se mais
simples, facilitando assim o processo de introspecção sobre o problema
proposto. Além do código ser mais eficiente, teremos também maior facilidade
de reutilização. No exemplo da \Cref{lst-nn} podemos observar como torna-se
simples utilizarmos métricas diferentes. No exemplo da \Cref{lst-dct} podemos
facilmente substituir a transformada ou alterar o tamanho dos blocos. Isto nos
garante agilidade e organização para analisar e comparar diferentes cenários do
nosso objeto de estudo.
  \subsection{Outras funções usuais em vetorização do código}

Diversas funções do Octave são uteis quando implementamos um código
utilizando a abordagem vetorial, permitindo a realização de operações complexas
em matrizes de maneira elegante e eficaz. Na \Cref{tbl-fnc-vec} listamos
algumas das principais funções que podem ser utilizadas para evitarmos \emph{loops}
explícitos, manipular, transformar, expandir ou classificar dados, tornando
seu código mais eficiente e legível. Algumas delas foram utilizadas nos
exemplos expostos neste trabalho, outras deixamos a cargo do leitor realizar a
leitura do \emph{help} da função e buscar outros exemplos de utilização na
internet.

\begin{table}[htpb]
\centering
\caption{Lista de funções usuais em vetorização do código.}\label{tbl-fnc-vec}
\begin{tabularx}{\linewidth}{lX}
\toprule
função & descrição \\
\midrule
\lstinline|all| & \small{determina se todos elementos no vetor são não nulos} \\
\lstinline|any| & \small{determina se ao menos um elemento no vetor é não nulo} \\
\lstinline|cat|, \lstinline|horzcat|, \lstinline|vertcat| & \small{concatenar matrizes} \\
\lstinline|deal| & \small{copiar parâmetros de entrada nos correspondentes parâmetros de saída} \\
\lstinline|diff| & \small{diferença móvel} \\
\lstinline|permute|, \lstinline|ipermute| & \small{permutar as dimensões de uma matriz} \\
\lstinline|prod| & \small{produtório} \\
\lstinline|repelems| & \small{replicar elementos de uma matriz} \\
\lstinline|repmat| & \small{replicar matriz} \\
\lstinline|reshape| & \small{rearranjar elementos da matriz} \\
\lstinline|shift| & \small{deslocar circularmente os elementos da matriz} \\
\lstinline|sort| & \small{ordenar os elementos da matriz} \\
\lstinline|sum|, \lstinline|cumsum|  & \small{soma e soma cumulativa dos elementos de uma matriz} \\
\lstinline|sub2ind|, \lstinline|ind2sub| & \small{conversão entre índice multidimensional e índice linear} \\
\lstinline|unique| & \small{retorna os elementos únicos} \\
\bottomrule
\end{tabularx}
\end{table}
 
\section{Conclusão}\label{sec-conclusao}

\epigraph{Eu não posso voltar para ontem porque eu era uma pessoa diferente
naquela época.\footnotemark} {Lewis Carroll, Alice no País das Maravilhas}
\footnotetext{Texto original: \emph{I can't go back to yesterday because I was
a different person then}.}

Neste artigo, exploramos a conexão entre notação e introspecção utilizando o
GNU Octave como ferramenta de programação para resolução de problemas. Através
da linguagem do Octave, notamos a ênfase dada à representação eficaz de
problemas matemáticos complexos. Além dos aspectos ligados à notação,
observamos como os recursos da linguagem e suas funções pré-definidas podem
levar a uma programação mais eficaz e eficiente. O domínio da notação vetorial
pode nos transformar, permitindo uma visão mais clara e direta dos problemas e
suas soluções.

O Octave, ao adotar uma notação próxima à álgebra linear, permite que os
programadores expressem suas intenções de forma explícita e concisa. Isso não
apenas facilita a codificação, mas também torna o código mais legível e
suscetível à introspecção. Através de recursos como a sobrecarga de operadores,
indexação, capacidade de lidar com arrays multidimensionais,
\emph{broadcasting} e \emph{function handles}, o Octave oferece uma plataforma
capaz de representar problemas matemáticos de maneira natural. Isso permite que
problemas sejam abordados de forma mais abstrata e flexível, alinhando-se à
visão de \textcite{iverson1980} de que a notação deve ser uma ferramenta que
amplie nossa capacidade de pensamento.

Em última análise, a abordagem através do Octave ilustra
vividamente como a notação não é apenas uma ferramenta para descrever
problemas, mas uma ferramenta para pensar sobre eles. Ao explorar essa conexão
entre notação e introspecção, esperamos que os programadores encontrem
inspiração para abordar desafios complexos de maneira mais eficaz, clara e
elegante. À medida que a programação evolui, a notação permanece como um pilar
fundamental, capacitando programadores a dar vida às suas visões e transformar
problemas em soluções.
 
\bibliographystyle{plain} \bibliography{article}